\documentclass[aps,twocolumn,floats,pre]{revtex4}
\usepackage{graphics,graphicx,epsfig}
\usepackage{amssymb,color}
\usepackage{epsf,epstopdf,wrapfig}

\newcommand{\beq}{\begin{equation}}
\newcommand{\eeq}{\end{equation}}
\newcommand{\beqn}{\begin{eqnarray}}
\newcommand{\eeqn}{\end{eqnarray}}

\begin{document}

\title{Optimizing information flow in small genetic networks. III. A self--interacting gene}

\author{Ga\v{s}per Tka\v{c}ik\footnote{gtkacik@ist.ac.at}$^{a,c}$, Aleksandra M. Walczak\footnote{awalczak@lpt.ens.fr}$^{b,c}$,
  and William Bialek\footnote{wbialek@princeton.edu}$^c$}

\affiliation{$^a$Institute of Science and Technology Austria, Am Campus 1, A-3400 Klosterneuburg, Austria\\
$^b$CNRS-Laboratoire de Physique Th\'eorique de l'\'Ecole Normale Sup\'erieure, 24 rue Lhomond, 75005 Paris, France\\
$^c$Joseph Henry Laboratories of Physics, Lewis--Sigler Institute for Integrative Genomics, Princeton University,
Princeton, New Jersey 08544, USA}

\date{\today}

\begin{abstract}
Living cells must control the reading out or ``expression'' of information encoded in their genomes, and this regulation often is mediated by transcription factors---proteins that bind to DNA and either enhance or repress the expression of nearby genes.  But the expression of transcription factor proteins is itself regulated, and many transcription factors regulate their own expression in addition to responding to other input signals.  Here  we analyze the simplest of such self--regulatory circuits, asking how parameters can be chosen to optimize information transmission from inputs to outputs in the steady state.  Some nonzero level of self--regulation is almost always optimal, with self--activation dominant when transcription factor concentrations are low and self--repression dominant when concentrations are high.  In steady--state the optimal self--activation is never strong enough to induce bistability, although there is a limit in which the optimal parameters are very close to the critical point.  
\end{abstract}

\maketitle

\section{Introduction}
In order to function and survive in the world, cells must make decisions about the reading out or ``expression'' of genetic information.  This happens when a bacterium makes more or less of an enzyme to exploit the variations in the availability of a particular type of sugar, and when individual cells in a multicellular organism commit to  particular fates during the course of embryonic development.  In all such cases, the control of gene expression involves the transmission of information from some input signal to the output levels of the proteins encoded by the regulated genes.   Although the notion of information transmission in these systems usually is left informal, the regulatory power that the system can achieve---the number of reliably distinguishable output states that can be accessed by varying the inputs---is measured, logarithmically, by the actual information transmitted, in bits \cite{shannon,inforeview}.  Since relevant molecules often are present at relatively low concentrations, or even small absolute numbers, there are irreducible physical sources of noise that will limit the capacity for information transmission.    Cells thus face a tradeoff between regulatory power (in bits) and resources (in molecule numbers).  What can cells do to maximize their regulatory power at fixed expenditure of resources?  More precisely, what can they do to maximize information transmission with bounded concentrations of the relevant molecules?  

We focus on the case of transcriptional regulation, where proteins---called transcription factors (TFs)---bind to sites along the DNA and modulate the rate at which nearby genes are transcribed into messenger RNA.  Because many of the regulated genes themselves code for TF proteins, regulatory interactions form a network.  The general problem of optimizing information flow through such regulatory networks is quite hard, and we have tried to break this problem into manageable pieces. Given the signal and noise characteristics of the regulatory interactions, cells can try to match the distribution of input transcription factor concentrations to these features of the regulatory network; even simple versions of this matching problem make experimentally testable predictions \cite{tkacik+al_08a,tkacik+al_08b}.  Assuming that this matching occurs, some regulatory networks still have more capacity to transmit information, and we can search for these optimal networks by varying both the topology of the network connections and the strengths of the interactions along each link in the network (the ``numbers on the arrows'' \cite{ronen+al_02}). We have addressed this problem first in simple networks where a single input transcription factor regulates multiple non--interacting genes \cite{tkacik+al_09}, and then in interacting networks where the interactions have a feedforward structure \cite{wt2010}.  But real genetic regulatory networks have loops, and our goal here is to study the simplest such case, where a single  input transcription factor controls a single self--interacting gene.  Does feedback increase the capacity of this system to transmit information?  Are self--activating or self--repressing genes more informative?  Since networks with feedback can exhibit multistability or oscillation, and hence a nontrivial phase diagram as a function of the underlying parameters, where in this phase diagram do we find the optimal networks?

Auto--regulation, both positive and negative, is one of the simplest and most commonly observed motifs in genetic regulatory networks \cite{motifs,keseler,hermsen}, and has been the focus of a number of experiments and modeling studies (see, for example, Refs \cite{bintu1,bintu2}).  A number of proposals have been advanced to explain its ubiquitous presence. Negative feedback (self--repression) can speed up the response of the genetic regulatory element \cite{rosenfeld}, and can reduce the steady state fluctuations in the output gene expression levels  \cite{becksei}.   Positive feedback (self--activation), on the other hand, slows down the dynamics of gene expression and sharpens the response of a regulated gene to its external input.  Self--activating genes could thus  threshold graded inputs,  transforming them into discrete, almost ``digital'' outputs \cite{becksei2}, allowing the cell  to implement binary logical functions \cite{buchler}. If self--activation is very strong, it can lead to multistability, or switch--like behavior of the response, so that  the genetic regulatory element can store the information for long periods of time \cite{lambda,billnips,gardner}; such elements will also exhibit hysteretic effects. Weak self--activation, which does not cause multistability, has been studied less extensively, but could play a role in allowing the cell to implement a richer set of input/output relations \cite{hermsen}.  Alternatively,  if the self--activating gene product can diffuse into neighboring nuclei of a multicellular organism, the sharpening effect of self--activation can compensate for the ``blurring'' of responses due to diffusion and hence open more possibilities for noise reduction through spatial averaging  \cite{bcd,erdmann}.

Many of the ideas about the functional role of auto--regulation are driven by considerations of noise reduction. The physical processes by which the regulatory molecules find and bind to their regulatory sites on the DNA, the operation of the transcriptional machinery, itself subject to thermal fluctuations, and the unavoidable shot noise inherent in producing a small number of output proteins all contribute towards  the stochastic nature of gene expression and thus place physical limits on the reliability of biological computation \cite{bialek+setayeshgar_05,kepler,thattai,sanchez_011}. In the past decade the advance of experimental techniques has enabled us to measure the total noise in gene expression and sometimes parse apart the contributions of various molecular processes towards this ``grand total'' \cite{elowitz+al_02,ozbudak+al_02,blake+al_03,raser+oshea_04,rosenfeld+al_05,pedraza+oudenaarden_05,ido_05,zenklusen_08,ido_11}.   With the detailed knowledge about noise in gene expression we can revisit the original question and ask: can both forms of auto-regulation  help mitigate the deleterious effects of noise on information flow through the regulatory networks and if so, how?

All of our  previous work in information transmission in transcriptional regulation has been in the steady state limit. A similar approach was taken in Ref \cite{ziv}, where the authors analyze information flow in elementary circuits including feedback, but with different model assumptions about network topology and noise. More  recently, de Ronde and colleagues have systematically reexamined the role of feedback regulation on the fidelity of signal transmission for time varying, Gaussian signals in cases where the (nonlinear) behavior of the genetic regulatory element can be linearized around some operating point \cite{deronde}. They found that auto--activation increases gain--to--noise ratios for low frequency signals, whereas auto--repression yields an improvement for high frequency signals.    While many of the functions of feedback involve dynamics, as far as we know all analyses of information transmission with dynamical signals resort to linear approximations.  Here we return to the steady state limit, where we can treat gene regulatory elements as fully nonlinear devices.  While we hope that our analysis of the self--regulated gene is interesting in itself, we emphasize that our goal is to build intuition for the analysis of more general networks with feedback.

\section{Formulating the problem}
\begin{figure}[t] 
\centering
\includegraphics[width=\linewidth]{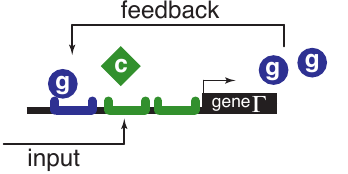} 
\caption{A schematic diagram of a self--regulating gene. The gene $\Gamma$ is depicted by a thick black line and a promoter start signal. Gene products $g$ denoted as blue circles can bind to the regulatory sites (one in this example) that control the expression of $\Gamma$. Direct control over the expression of $\Gamma$ is exerted by molecules of the transcription factor  $c$ (green diamonds, two binding sites). } 
\label{f1}
\end{figure}

Figure \ref{f1}  shows a schematic of the system that we will analyze in this paper, a gene $\Gamma$ that is controlled by two regulators: directly by an external transcription factor, as well as in a feedback fashion by its own gene products. We will refer to the transcription factor as the regulatory \emph{input}; its concentration in the relevant (cellular or nuclear) volume $\Omega$ will be denoted by $c$. In addition, the gene products of $\Gamma$, whose number in the relevant volume $\Omega$ we denote by $G$ and to which we refer to as the \emph{output}, can also bind to the regulatory region of $\Gamma$, thereby activating or repressing the gene's expression.   As we attempt to make our description of this system mathematically precise, the heart of our model will be the   \emph{regulatory function} that maps   the concentrations of the two regulators at the promotor region of $\Gamma$ to the rate at which output molecules are synthesized. 

We can write the equation for the dynamics of gene expression from $\Gamma$ by assuming that synthesis and degradation of the gene products are single kinetic steps, in which case we have
\begin{equation}
\frac{dG}{dt}=r_{\rm max} f(c,\gamma) - \frac{1}{\tau}G + \xi; \label{eqd}
\end{equation}
here, $r_{\rm max}$ is the maximum rate for production of $G$,  $\tau$ is the protein degradation time, and $\gamma$ is the concentration of the output molecules in the relevant volume $\Omega$.   To include the noise effects inherent in creating and degrading single molecules of $G$ we introduce the   Langevin   force $\xi$, and we will discuss the nature of this and other noise sources in detail later.  Importantly, departures from our simplifying assumptions about the kinetics can, in part, be captured by proper treatment of the noise terms, as discussed below.

We are interested in the information that the steady state output of $\Gamma$ provides about the input concentration $c$.  Following our previous work \cite{tkacik+al_09,wt2010}, we address this problem in stages. First we relate information transmission to the response properties and noise in the regulatory element, using a small noise approximation to allow analytic progress (Section \ref{infsectA}).  Then we show how the  relevant noise variances can be computed from the model in Eq (\ref{eqd}), taking advantage of our  understanding of the physics underlying the essential sources of noise (Section \ref{insectB}); this discussion is still quite general, independent of the details of the regulation function.  Then we explain our choice of the regulation function, adapted from the Monod--Wyman--Changeux description of allosteric interactions (Section \ref{MWC}).  Because feedback allows for bifurcations, we have to map the phase diagram of our model (Section \ref{phasediag}), and develop approximations for the information transmission near the critical point (Section \ref{critinfo}) and in the bistable regime (Section \ref{infobi}).    Our discussion reviews some earlier results, in the interest of being self--contained, but the issues in Sections \ref{phasediag}--\ref{infobi} are all new to the case of networks with feedback.

\subsection{Noise and information transmission}
\label{infsectA}

We are interested in computing the mutual information between the input and the output of a regulatory element, in steady state.  We have agreed that the input signal is the concentration $c$ of the transcription factor, and we will take the output to be the concentration $g$ of the gene products, which we colloquially call the expression level of the gene $\Gamma$.  An important feature of the information transmission is that its mathematical definition is independent of the units that we use in measuring these concentrations, so when we later choose some natural set of units we won't have to worry about substituting into the formulae we derive here.

Following Shannon \cite{shannon}, the mutual information between $c$ and $g$ is defined by 
\begin{equation}
I(c;g) = \int dc\int dg\, P(c,g) \log_2 \left[  {{P(c,g)}\over{P_{\rm in}(c) P_{\rm out}(g)}}\right] \, {\rm bits},
\label{Isym}
\end{equation}
where input concentrations $c$ are drawn from the distribution $P_{\rm in}(c)$, the output expression levels that we can observe are drawn from the distribution $P_{\rm out}(g)$, and the joint distribution of these two quantities is $P(c,g)$.  We think of the expression level as responding to the inputs, but this response will be noisy, so given the input $c$ there is a conditional distribution $P(g|c)$.  Then the symmetric expression for the mutual information in Eq (\ref{Isym}) can be rewritten as a difference of entropies,
\begin{equation}
I(c;g)=S[P_{\rm out}(g)] - \int dc\; P_{\rm in}(c)  S[P(g|c)] ,
\label{info_ent}
\end{equation}
where the entropy of a distribution is defined, as usual, by
\begin{equation}
S[P(x)] = -\int dx\, P(x) \log_2 P(x) .
\end{equation}
Finally, we recall that 
\begin{equation}
P_{\rm out}(g) = \int dc\; P_{\rm in}(c) P(g|c) .
\label{pg}
\end{equation}

Notice that the mutual information is a functional of two probability distributions, $P_{\rm in}(c)$ and $P(g|c)$.  The latter distribution describes the response and noise characteristics of the regulatory element, and is something we will be able to calculate from Eq (\ref{eqd}).   Following Refs~\cite{tkacik+al_08a,tkacik+al_08b,tkacik+al_09, wt2010}, we may then ask: given that $P(g|c)$ is determined by the biophysical properties of the genetic regulatory element, what is the optimal choice of $P_{\rm in}(c)$ that will maximize the mutual information $I(c;g)$? To this end we have to solve the problem of extremizing 
\begin{equation}
\mathcal{L}[P_{\rm in}(c)] = I(c;g) - \Lambda\int dc\; P_{\rm in} (c),
\label{extremal}
\end{equation}
where the Lagrange multiplier $\Lambda$ enforces the normalization of $P_{\rm in}(c)$. Other ``cost''  terms are possible, such as adding a term proportional to $\int dc\,cP_{\rm in}(c)$, which would penalize the average cost of input molecules $c$, although here we take the simpler approach of fixing the maximum possible value of $c$, which is almost equivalent \cite{tkacik+al_09}.
If the noise were truly zero, we could write the distribution of outputs as 
\begin{equation}
P_{\rm out}(g) = \int dc\, P_{\rm in}(c) \delta[ g - \bar g (c)] ,
\end{equation}
where $\bar g (c)$ is  the average output as a function of the input, i.e. the mean of the distribution $P(g|c)$.  Then if the function $\bar g (c)$ is invertible, we can write the entropy of the output distribution as
\begin{eqnarray}
S[P_{\rm out}(g)] &\equiv& -\int dg\, P_{\rm out}(g)\log_2 P_{\rm out}(g)\nonumber\\
&\rightarrow& -\int dc\, P_{\rm in}(c) \log_2\left[ P_{\rm in}(c) {\bigg |} {{d\bar g (c)}\over {dc}}{\bigg |^{-1}}\right] ,
\end{eqnarray}
and we can think of this as the first term in an expansion in powers of the noise level \cite{tkacik+al_08a}.  Keeping only this leading term, we have
\begin{widetext}
\begin{equation}
\mathcal{L}[P_{\rm in} (c)] = -\int dc\, P_{\rm in}(c) \log_2\left[ P_{\rm in}(c) {\bigg |} {{d\bar g (c)}\over {dc}}{\bigg |^{-1} }\right]  -  \int dc\, P_{\rm in}(c)  S[P(g|c)] - \Lambda\int dc\; P_{\rm in} (c), \label{pointback}
\end{equation}
\end{widetext}
and one can then show that the extremum of $\mathcal{L}$ occurs at
\begin{equation}
P_{\rm in}^*(c)=\frac{1}{Z} 2^{-S[P(g|c)]}\left| \frac{d\bar{g}(c)}{dc}\right|, \label{gensol}
\end{equation}
where the entropy is measured in bits, as above, and the normalization constant 
\begin{equation}
Z = \int dc \, \left| \frac{d\bar{g}(c)}{dc}\right| 2^{-S[P(g|c)]} .
\end{equation}
The maximal value of the mutual information is then simply $I^* = \log_2 Z$.

In the case where $P(g|c)$ is Gaussian, 
\begin{equation}
P(g|c) =  {1\over\sqrt{2\pi \sigma_g^2(c)}} \exp\left[ - {{(g-\bar g (c))^2}\over{2\sigma_g^2 (c)}}\right] ,
\end{equation}
the entropy is determined only by the variance $\sigma_g^2 (c)$,
\begin{equation}
S[P(g|c)]={1\over 2} \log_2\left[ 2\pi e \sigma_g^2(c)\right] .
\end{equation}
It is useful to think about propagating this (output) noise variance back through the input/output relation $\bar g (c)$, to define the effective noise at the input,
\begin{equation}
\sigma_c^2(c)=\left| \frac{d\bar{g}(c)}{dc} \right|^{-2}\sigma_g^2 . \label{sigmac}
\end{equation}
Then we can write
\begin{equation}
P_{\rm in} ^*(c)=\frac{1}{Z} \frac{1}{\sqrt{2\pi e}\sigma_c(c)} . \label{optpc}
\end{equation}
As before, $Z$ is the normalization constant,
\begin{equation}
Z=\int_0^Cdc\; \left[\frac{1}{2\pi e}\left(\frac{d\bar{g}}{dc}\right)^2 \frac{1}{\sigma^2_g(c)} \right]^{1/2}, \label{eq2}
\end{equation}
where $C$ is the maximal value of the input concentration, and again we have the information  $I^*(c;g) = \log_2Z \,{\rm  bits}$. 

\subsection{Noise variances}
\label{insectB}

Equation (\ref{eq2}) relates $Z$, and hence the information transmission $I^* = \log_2 Z$, to the steady state response and noise in our simple regulatory element.  These quantities are calculable from the dynamical model in Eq (\ref{eqd}), if we understand the sources of noise.    There are two very different kinds of noise that we need to include in our analysis.

First, we are describing molecular events that synthesize and degrade individual molecules, and individual molecules behave randomly.  If we say that there is synthesis of $\bar r$ molecules per second on average, then if the synthesis is limited by a single kinetic step, and if all molecules behave independently, then the actual rate $r(t)$ will fluctuate with a correlation function $\langle \delta r (t) \delta r(t')\rangle = \bar r \delta(t-t')$.  Similarly, if on average there is degradation of $\bar d$ molecules per second, then the actual degradation rate $d(t)$ will fluctuate with $\langle \delta d (t) \delta d(t')\rangle = \bar d \delta(t-t')$.  Thus if we want to describe the time dependence of the number of molecules $N(t)$, we can write
\begin{equation}
{{dN}\over{dt}} = r(t) - d(t) = \bar r - \bar d + \xi (t) ,
\end{equation}
where 
\begin{equation}
\xi(t) = \delta r(t) - \delta d(t) .
\end{equation}
If we are close to the steady state,  $\bar r = \bar d$, and if synthesis and degradation reactions are independent, we have
\begin{equation}
\langle \xi(t) \xi(t') \rangle = 2 \bar d \delta(t-t') .
\end{equation}
If some of the reactions involve multiple kinetic steps, or if the molecules we are counting are amplified copies of some other molecules, then the noise will be proportionally larger or smaller, and we can take account of this by introducing a ``Fano factor'' $\nu$, so that
\begin{equation}
\langle \xi(t) \xi(t') \rangle \rightarrow 2 \nu \bar d \delta(t-t') .
\end{equation}
For more about the Langevin description of noise in chemical kinetics, see Ref \cite{gardiner}. 

The second irreducible source of noise is that the synthesis reactions are regulated by transcription factor binding to DNA, and these molecules arrive randomly at their targets.  One way to think about this is that the concentrations of TFs which govern the synthesis rate are not the bulk average concentrations over the whole cell or nucleus, but rather concentrations in some small ``sensitive volume'' determined by the linear size $\ell$  of the targets themselves \cite{bialek+setayeshgar_05,tkacik+bialek_07,tkacik+al_08c}.  Concretely, if we write the synthesis rate as 
\begin{equation}
r = r_{\rm max} f(\hat c,\gamma) ,
\end{equation}  
where $\hat c$ is the local concentration of the input transcription factor and $\gamma$ is the concentration of the gene product that feeds back to regulate itself, we should really think of these concentrations as
$\hat c = c+\xi_c$ and $\gamma = G/\Omega + \xi_g$, where we separate the mean values and the local fluctuations; note that the mean gene product concentration is the ratio of the molecule number $G$ to the relevant volume $\Omega$.  The local concentration fluctuations are also white, and the spectral densities are given accurately by dimensional analysis  \cite{bialek+setayeshgar_05,tkacik+bialek_07}, so that
\begin{eqnarray}
\langle \xi_c (t) \xi_c(t') \rangle  &=& (2 c /D\ell) \delta (t-t') \\
\langle \xi_\gamma (t) \xi_\gamma (t') \rangle  &=& (2 G /\Omega D\ell) \delta (t-t') ,
\end{eqnarray}
where $D$ is the diffusion constant of the transcription factor molecules, which we assume is the same for the input and output proteins.

We can put all of these factors together if the noise is small, so that  it drives fluctuations which stay in the linear regime of the dynamics.  Then  if the steady state solution to Eq (\ref{eqd}) in the absence of noise is denoted by $G= \bar G(c)$,  we can linearize in the fluctuations $\delta G = G - \bar G$:
\begin{widetext}
\begin{eqnarray}
\frac{dG}{dt}&=&r_{\rm max} f(\hat c ,\gamma) - \frac{1}{\tau}G + \xi\nonumber\\
&=& r_{\rm max} f( c + \xi_c , G/\Omega + \xi_\gamma ) - \frac{1}{\tau}G + \xi \label{LL1}\\
\Rightarrow \frac{d(\delta G)}{dt} &=& \left[ {{r_{\rm max}}\over\Omega} {{\partial f(c,\gamma)}\over{\partial \gamma}}{\bigg |}_{\gamma = \bar G /\Omega} - {1\over\tau}\right] \delta G + \xi_{\rm eff}(t) ,\,{\rm where} \label{LL2}\\
\langle \xi_{\rm eff}(t) \xi_{\rm eff}(t')\rangle &=& 2 \left[ \nu \frac{\bar G}{\tau} +  \left( r_{\rm max} {{\partial f(c,\gamma)}\over{\partial \gamma}}{\bigg |}_{\gamma = \bar G /\Omega}\right)^2  \frac{2\bar G}{\Omega D \ell}
+  \left( r_{\rm max} {{\partial f(c,\bar G /\Omega)}\over{\partial c}}\right)^2 \frac{2c}{ D \ell} \right] \delta(t-t') \label{LL3}.
\end{eqnarray}

To solve this problem and compute the variance in the output number of molecules $\langle (\delta G)^2\rangle$, it is useful to recall the Langevin equation for  the position $x(t)$ of an overdamped mass tethered by a spring of stiffness $\kappa$, subject to a drag force proportional to the velocity, $F_{\rm drag} = - \alpha v$:
\begin{eqnarray}
\alpha \frac{dx}{dt} &=& -\kappa x + \zeta(t)\\
\langle \zeta (t)\zeta (t') \rangle &=& 2 k_B T \alpha \delta(t-t') .
\end{eqnarray}
From equipartition we know that these dynamics predict the variance $\langle x^2\rangle = k_B T/\kappa$. Identifying terms with our Langevin description of the synthesis and degradation reactions, we find
\begin{equation}
\langle (\delta G )^2 \rangle = {1\over{1\over\tau} -  {{r_{\rm max}}\over\Omega} {{\partial f }\over{\partial \gamma}}} \left[ \nu \frac{\bar G}{\tau} +  \left( r_{\rm max} {{\partial f}\over{\partial \gamma}} \right)^2  \frac{\bar G}{\Omega D \ell} +  \left( r_{\rm max} {{\partial f}\over{\partial c}}\right)^2 \frac{c}{ D \ell} \right]  ,
\end{equation}
where we understand that the partial derivatives of $f$ are to be evaluated at the steady state $\gamma = \bar G/\Omega$.  

We have defined $r_{\rm max}$ as the maximum synthesis rate, so that  the regulation function $f$ is in the range $0 \leq f \leq 1$, and hence the maximum mean expression level is $\bar G_{\rm max} = r_{\rm max}\tau$.  Thus it makes sense to work with a normalized expression level $g\equiv G/(r_{\rm max}\tau)$, and to think of the regulation function $f$ as depending on $g$ rather than on the absolute concentration $\gamma$.  Then we have 
\begin{equation}
\langle (\delta g )^2 \rangle \equiv \sigma_g^2 (c)= {1\over{1  -  (\partial f/\partial g)}}  \left[ \left( \frac{\nu}{r_{\rm max}\tau} \right) \bar{g}  +  \left(   {{\partial f}\over{\partial g}} \right)^2  \bar{g} \frac{1}{D \ell\gamma_{\rm max}} +  \left( {{\partial f}\over{\partial c}}\right)^2 \frac{c}{ D \ell} \right]  ,
\end{equation}
where $\gamma_{\rm max}$ is the maximal mean concentration of output molecules.  As discussed previously \cite{tkacik+al_09}, we can think of $N_g =r_{\rm max}\tau/\nu$ as the maximum number of independent output molecules, and this combines with the other parameters in the problem to define a natural concentration scale, $c_0 = N_g/(D\ell\tau)$.  Once we choose units where $c_0 = 1$, we have a simpler expression,
\begin{equation}
\sigma_g^2 (c)  = {1\over {N_g}} \cdot {1\over{1  -  (\partial f/\partial g)}}  \left[  \bar{g}  +  \left(   {{\partial f}\over{\partial g}} \right)^2   \frac{\bar{g}}{\gamma_{\rm max}} +  \left( {{\partial f}\over{\partial c}}\right)^2  c  \right]  ,
\end{equation}
where we notice that almost all the parameters have been eliminated by our choice of units.

Finally, we need to use the variance to compute the information capacity of the system, $I^* = \log_2 Z$, where  from Eq (\ref{eq2}) we have
\begin{eqnarray}
Z&=&\int_0^Cdc\; \left[\frac{1}{2\pi e}\left(\frac{d\bar{g}}{dc}\right)^2 \frac{1}{\sigma^2_g(c)} \right]^{1/2}  =  \left[ \frac{N_g}{2\pi e}\right]^{1/2}\tilde Z,\\
\tilde Z &=&  \int_0^Cdc\;\left[ \left(\frac{d\bar{g}}{dc}\right)^2 \frac{1  -  (\partial f/\partial g)}{ \bar{g}  +  (   \partial f/\partial g )^2   (\bar{g}/\gamma_{\rm max}) +  (\partial f/ \partial c )^2  c   } \right]^{1/2} ; 
\end{eqnarray}
\end{widetext}
$C$ is the maximum concentration of input transcription factor molecules, in units of $c_0$. Notice that the parameter $N_g$ just scales the noise and (in the small noise approximation) thus adds to the information, $I^* = \log_2 \tilde Z + (1/2)\log_2(N_g/2\pi e)$; the problem of optimizing information transmission thus is the problem of optimizing $\tilde Z$.  Further, because 
\begin{equation}
\bar{g} = f(c,\bar{g}), \label{ss}
\end{equation}
the total derivative $d\bar g /dc$ can be expressed though
\begin{equation}
\frac{d\bar{g}}{dc}=\frac{\partial f}{\partial c}  +  \frac{\partial f}{\partial g}\frac{d\bar{g}}{dc}.
\end{equation}
Putting all of these pieces together, we find
\begin{equation}
\tilde Z= \int_0^C dc \frac{\frac{\partial f}{\partial c}\left(1-\frac{\partial f}{\partial g}\right)^{-\frac{1}{2}}}{\sqrt{ \bar{g} + \left(\frac{\partial f}{\partial c}\right)^2c + \left(\frac{\partial f}{\partial g}\right)^2\bar{g}/\gamma_{\rm max}}}. \label{cap1}
\end{equation}
In what follows we will start with the assumption that, since both input and output molecules are transcription factor proteins, their maximal concentrations are the same, and hence $\gamma_{\rm max} = C$; we will return to this assumption at the end of our discussion.

If a regulatory function $f(c,g)$ is chosen from some parametric family,  Eq~(\ref{cap1}) allows us to compute the information transmission as a function of these parameters and search for an optimum.  Before embarking on this path, however, we note that the integrand of $\tilde Z$ can have a divergence if $\partial f/\partial g = 1$. This  is a condition for the existence of a \emph{critical point}, and in this simple system the critical point or bifurcation separates the regime of monostability from the regime of bistability.    We expect that at this point the fluctuations around $\bar{g}$ are no longer Gaussian, and we need to compute higher order moments. Thus, Eq (\ref{cap1}), as is, can safely be used only in the monostable regime away from the critical point; in Section \ref{critinfo} we compute the expression for the mutual information near to and at the critical point for a particular choice of $f$.  There are even more problems in the bistable regime, since there are multiple solutions to Eq (\ref{ss}), and  in Section \ref{infobi} we discuss information in the bistable regime.

\subsection{MWC regulatory function}
\label{MWC}

To continue, we must choose a regulatory function. In Ref~\cite{wt2010}, where we analyzed genetic networks with feedforward interactions, we studied Hill--type regulation \cite{hill} and Monod--Wyman--Changeaux--like (MWC) regulation \cite{mwc}, and found that the MWC family encompasses a broader set of functions than the Hill family; for a related discussion see  Ref \cite{mirny}. MWC functions also allow for a natural introduction of convergent control, where a node in a network is simultaneously regulated by several types of regulatory molecules. Briefly, in the MWC model one assumes that the molecule or supermolecular complex being considered has two states, which we identify here with ON and OFF states of the promoter.  The binding of each different regulatory factor is always independent, but the binding energies depend on whether the complex is in an OFF or ON state, so that (by detailed balance) binding shifts the equilibrium between these two states.   

In our case, we have two regulatory molecules, the input transcription factor with concentration $c$ and the gene product with concentration $g$.  If there are, respectively, $n_c$ and $n_g$ identical binding sites for these molecules then the probability of being in the ON state is
\begin{widetext}
\begin{equation}
f(c,g) = {{(1+c/Q_c^{\rm on})^{n_c}(1+g/Q_g^{\rm on})^{n_g}} \over{L (1+c/Q_c^{\rm off})^{n_c}(1+g/Q_g^{\rm off})^{n_g} + (1+c/Q_c^{\rm on})^{n_c}(1+g/Q_g^{\rm on})^{n_g}}} ,
\end{equation}
\end{widetext}
where $Q_c^{\rm on} , Q_g^{\rm on}$ are the binding constants in the ON state, and similarly $Q_c^{\rm off} , Q_g^{\rm off}$ are the binding constants in the OFF state; $L$ reflects the ``bare'' free energy difference between the two states.    If the binding of the regulatory molecules has a strong activating effect, then we expect $Q_c^{\rm on} \ll Q_c^{\rm off}$, and similarly for $Q_g$, which means that only one binding constant is relevant for each molecule, and we will refer to these as $K_c$ and $K_g$.  Then  we can write
\begin{eqnarray}
f(c,g) &=& {1\over{1 + e^{-F(c,g)}}}\label{mwc}\\
F(c,g) &=&  n_c \ln\left( {{1 + c/K_c}\over{1+c_{1/2}/K_c}}\right)  + n_g \ln\left( {{1 + g/K_g}\over{1+1/(2K_g)}}\right) ,\nonumber\\
&&\label{F_def}
\end{eqnarray}
where $c_{1/2}$ is the input concentration at which  $\bar g(c_{1/2}) = 1/2$.  Notice that if binding of $g$ strongly represses the gene, then we have $Q_g^{\rm off} \ll Q_g^{\rm on}$, but this can be simulated by changing the sign of $n_g$.  Thus we should think of the parameters $n_c$ and $n_g$ as being not just the number of binding sites, but also an index of activation vs. repression.  We will also treat these parameters as continuous, which is  a bit counterintuitive but allows us to describe, approximately, situations in which the multiple binding sites are inequivalent, or in which $Q^{\rm on}$ and $Q^{\rm off}$ are not infinitely different.

From the discussion in the previous section, we will need to evaluate the partial derivatives of $f(c,g)$ with respect to it arguments.  For the MWC model, these derivatives take simple forms:
\begin{eqnarray}
\frac{\partial f}{\partial c} &=& f(1-f) \frac{n_c}{K_c+c}		\\
\frac{\partial f}{\partial g} &=& f(1-f) \frac{n_g}{K_g+g}. \label{g_derivative}
\end{eqnarray}
\subsection{Phase diagram}
\label{phasediag}
\begin{figure}[t] 
\centering
\includegraphics[width=2.0in]{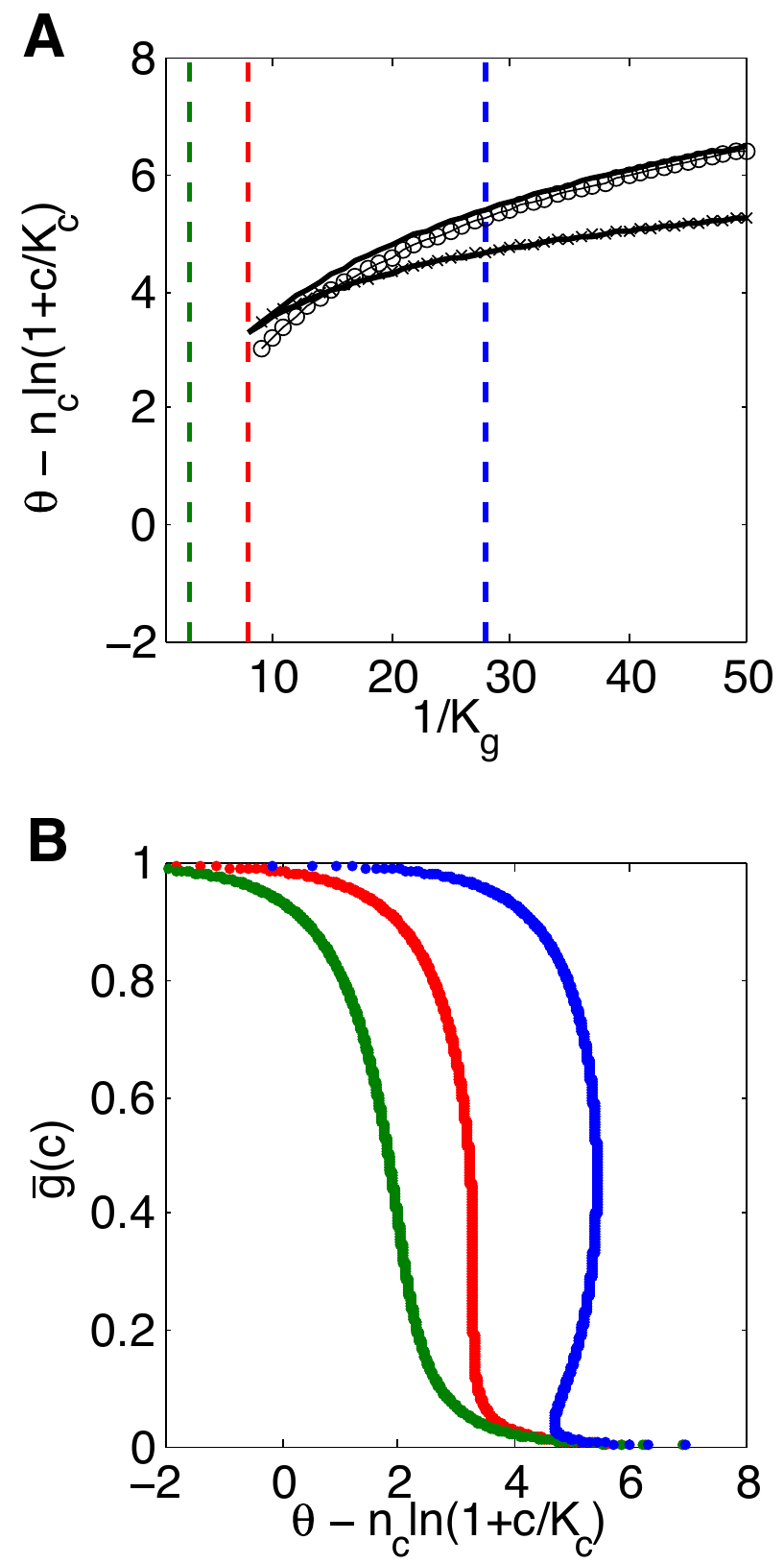} 
\caption{Monostable (green), critical (red) and bistable (blue) behavior of the self-activating gene for $n_g=2$. {\bf A)} The phase diagram as a function of $K_g$ and the input-dependent term. In the region between the black solid lines two solutions $\bar{g}_{1,2}$ exist for every value of the input $c$ (y-axis). The corresponding critical value is at $K_g^*=1/8$  (cusp of the black solid lines). Circles and crosses represent analytical approximations to the exact boundary of the bistable region for $n_g=2$ and large $1/K_g$ (see text). For three choices of $K_g$ denoted by vertical dashed lines, the input / output relations $\bar{g}(c)$ are plotted in B. {\bf B)}. The critical solution (red) has an infinite total derivative $\bar{g}'(c)$ at $\theta-n_c\log(1+c/K_c)=3.29$, $\bar{g} = 0.25$. The bistable system (blue) has three solutions, two stable and one unstable, for a range of inputs that can be read out from the plot in A. } 
\label{f-bistableregion}
\end{figure}

Let us start by examining the stability properties of Eq (\ref{ss}), which determines the steady state $\bar g(c)$.  Viewed as a function of $g$, $f(c,g)$ is sigmoidal, and so if we try to solve $g=f(c,g)$ graphically we  are  looking for the intersection of a sigmoid  with the diagonal, as a function of $g$.  In doing this we expect that, for some values of the parameters, there will be exactly one solution, but that as we change parameters (or the input $c$), there will be a transition to multiple solutions. This transition happens when $f$ just touches the diagonal, that is, when for some $\bar{g}^*$ it holds true that $f(\bar{g}^*,c)=\bar{g}^*$ and $\partial f(\bar{g},c)|_{\bar{g}^*}/\partial \bar{g}=1$. Using Eq (\ref{g_derivative}), these two conditions can be combined to yield an equation for $\bar{g}^*$:
\begin{equation}
\bar{g}^*(1-\bar{g}^*)\frac{n_g}{K_g+\bar{g}^*}=1 \label{ccrit}
\end{equation}
This is a quadratic in $\bar{g}^*$ for which no real solution on $\bar{g}^*\in[0,1]$ exists if either
\begin{equation}
K_g > \frac{(n_g-1)^2}{4n_g}
\end{equation}
or $n_g<1$. When either of these conditions are fulfilled, the gene is in the monostable regime. At the critical point, $K_g^*=(n_g-1)^2/(4n_g)$ and $\bar{g}^*=(n_g-1)/(2n_g)$. This is illustrated in Fig~\ref{f-bistableregion}, as a function of the effective input  $\chi(c)=\theta-n_c\ln(1+c/K_c)$, where, from Eq (\ref{F_def}),
$\theta = n_c \ln(1+c_{1/2}/K_c) + n_g \ln(1+1/(2K_g))$.

For the special case of $n_g=2$ it is not hard to compute the analytical approximations for the boundary of the bistable domain. First, Eq~(\ref{mwc}) can be expanded for large $K_g$, yielding a quadratic equation for $\bar{g}$ that has two solutions only when
\begin{equation}
\chi < -\ln 4 - 2\ln K_g.
\end{equation}
To get the lower bound, we expand Eq~(\ref{mwc}) for small $\bar{g}$ and retain terms up to the quadratic order in $\bar{g}$; the resulting quadratic equation yields two solutions only if 
\begin{equation}
\chi > \ln(4/K_g - 3).
\end{equation}
Both approximations are plotted as circles and crosses, respectively, in Fig~\ref{f-bistableregion}, and match the exact curves well. For other values of $n_g$ we solve Eq~(\ref{ss}) exactly, using a bisection method to get all solutions for a given $c$ and we partition the range of $c$ adaptively into a denser grid where the derivative $\bar{g}'(c)$ is large. For integer values of $n_g$ when the equation can be rewritten as a polynomial in $\bar{g}$, it is technically easier to find the roots of the polynomial; alternatively one can solve for $c$ given $\bar{g}$ using a simple bisection, because $c(\bar{g})$ is an injective function. 
\subsection{Information transmission near  the critical point}
\label{critinfo}
In this section we will generalize the computation of noise and information in the region close to the critical point, where the Gaussian noise approximation breaks down.   We start by rewriting Eq's (\ref{LL1}--\ref{LL2}) in our normalized units,
\begin{eqnarray}
\tau {{dg}\over{dt}} &=& f(c,g) - g + \zeta\\
\langle \zeta(t)\zeta(t') \rangle &=& 2\tau T(g)\delta (t-t')\\
T(g) &=& {1\over {N_g}} \left[ g + \left( {{\partial f}\over{\partial g}}\right)^2 {g\over{\gamma_{\rm max}}} + \left( {{\partial f}\over{\partial c}}\right)^2 c\right] .
\end{eqnarray}
This is equivalent to Brownian motion of the coordinate $g$ in a potential $V(g,c)$ defined by
\begin{equation}
- {{dV(g,c)}\over{dg}} = f(c,g) - g ,
\end{equation}
with an effective  temperature $T(g)$ that varies with position.  If we simulate this Langevin equation, we will draw samples out of the distribution $P(g|c)$,  but we can construct this distribution directly by solving the equivalent diffusion or Fokker--Planck equation,
\begin{equation}
\frac{\partial P(g,t)}{\partial t}=\frac{\partial}{\partial g}  \left[V'(g,c)P(g,t)\right]+ \frac{\partial^2} {\partial g^2}\left[T(g)P(g,t)\right];
\end{equation}
the steady--state solution is then $P(g|c)$, and this is
\begin{equation}
P(g|c) = P(0) \frac{T(0)}{T(g)}\exp\left[-\int_0^g \frac{V'(y,c)}{T(y)}dy\right]. \label{intg}
\end{equation}

The ``small noise approximation'' in this extended framework corresponds to expanding the integrand in Eq~(\ref{intg}) around the mean, $g=\bar{g}(c)+\delta g$. If we write $x = g - \bar g(c)$, we will find
\begin{equation}
P(g|c) \propto \exp\left( -a_2 x^2-a_3 x^3-a_4 x^4 - \dots \right), \label{pquartic}
\end{equation}
where our previous approximations correspond to keeping only $a_2$.  The critical point is where $a_2 = 0$, and we have to keep higher order terms.  In principle the expansion coefficients have contributions from the $g$--dependence of the effective temperature, but we have checked that these contributions are negligible near criticality.  Then we have
\begin{eqnarray}
a_2 &=& \frac{1}{2!}\left[\frac{1-f'}{T}\right] ,\label{eqset}\\
a_3 &=& -\frac{1}{3!} \frac{f''}{T}, \nonumber\\
a_4 &=& -\frac{1}{4!}\frac{f'''}{T} ,\nonumber
\end{eqnarray}
where primes denote derivatives with respect to $g$, and all terms should be evaluated at $g = \bar g(c)$.

\begin{widetext}

For the Monod--Wyman--Changeaux regulatory function in Eq (\ref{mwc}), all these derivatives can be evaluated explicitly:
\begin{eqnarray}
f' = \frac{\partial f}{\partial g}&=&f(1-f)\frac{n_g/K_g}{1+g/K_g}	\label{dfdg1}\\
f'' = \frac{\partial^2 f}{\partial g^2}&=&f(1-f)\left(\frac{n_g/K_g}{1+g/K_g}\right)^2\left(1-2f-\frac{1}{n_g}\right) \label{dfdg2} \\
f''' = \frac{\partial^3f}{\partial g^3}&=&f(1-f)\left(\frac{n_g/K_g}{1+g/K_g}\right)^3\left[\left(1-2f-\frac{1}{n_g}\right)\left(1-2f-\frac{2}{n_g}\right)-2f(1-f)\right] \label{dfdg3}
\end{eqnarray}
\end{widetext}
From Eq~(\ref{ccrit}), the critical point occurs at $\bar{g}^*=(n_g-1)/2n_g$ when $K_g^*=(n_g-1)^2/4n_g$, and at this point the derivatives simplify:
\begin{eqnarray}
f' &=&1 \\
f'' &=& 0	\\
f''' &=&-\frac{8 n_g^2}{n_g^2-1}. 
\end{eqnarray}

Now we want to explore behavior in the vicinity of the critical point; we will fix $K_g$ to its critical value, $K_g^*(n_g)$, and compute the derivatives in Eqs~(\ref{dfdg1}-\ref{dfdg3}) as $\partial f/\partial g\rightarrow 1$.  Consider therefore a small positive $\epsilon$ such that
\begin{equation}
\frac{\partial f}{\partial g}\bigg|_{\bar{g}}=1-\epsilon.
\end{equation}
In a system with chosen $K_g$ and $n_g$ that yield critical behavior, the deviation from criticality above will happen at $\bar{g}=\bar{g}^*+\Delta$. To find the relation between $\epsilon$ and $\Delta$, we evaluate the derivative in Eq (\ref{dfdg1}) at $\bar{g}$ to form a function $\psi(\bar{g})=\bar{g}(1-\bar{g}) n_g/(K_g+\bar{g})$, which evaluates to 1 at $\bar{g}^*$. This function can be expanded in Taylor series around $\bar{g}^*$; the first order in $\Delta$ vanishes and we find:
\begin{equation}
\frac{n_g(K_g+1)}{(1+\bar{g}^*/K_g)^3}\Delta^2=\epsilon.
\end{equation}
Therefore, the derivative deviates by $\epsilon$ from criticality at 1 when $\bar{g}$ deviates by $\pm \Delta$ from the $\bar{g}^*$. We now perform similar expansions on the second- and third-order derivatives in Eqs (\ref{dfdg2},\ref{dfdg3}), and evaluate the factors at the critical point:
\begin{eqnarray}
\frac{\partial f}{\partial g}&=&1 - \frac{4n_g^2}{n_g^2-1}\Delta^2 \label{perturb1}\\
\frac{\partial^2 f}{\partial g^2}&=& -\frac{8n_g^2}{n_g^2-1}\Delta+\frac{32 n_g^3}{(n_g^2-1)^2}\Delta^2	\label{perturb2}\\
\frac{\partial^3 f}{\partial g^3}&=&-\frac{8 n_g^2}{n_g^2-1} + \frac{64 n_g^3}{(n_g^2-1)^2}\Delta + \label{perturb3} \\
&+&\frac{128 n_g^4(n_g^2-4)}{(n_g^2-1)^3}\Delta^2. \nonumber
\end{eqnarray}
These expressions have been evaluated for $K_g^*$, but we could have easily repeated the calculation by assuming that $K_g$ itself can deviate a bit from the critical value, i.e. $K_g=K_g^*+\delta K_g$, which would yield somewhat more complicated results that we don't reproduce here.  

Equations~(\ref{perturb1}--\ref{perturb3}) can  be used in Eq (\ref{eqset}) to write down the probability distribution $P(g|c)$. Far away from the critical point the Gaussian approximation is assumed to hold, and $a_3,a_4$ can be set to 0. Close to the critical point the higher order terms $a_3$ and $a_4$ need to be included. To assess the range where this switchover occurs, we compare in Eqs (\ref{perturb1}--\ref{perturb3}) the leading to the subdominant correction: we insist that the quadratic correction in Eq~(\ref{perturb2}) is always smaller than linear, and that the linear correction in Eq (\ref{perturb3}) is always smaller than constant (we drop the quadratic correction there). We found empirically that including the higher-order corrections yields good results when the following conditions are simultaneously satisfied:
\begin{eqnarray}
|\Delta| &<& \frac{n_g^2-1}{16n_g} \label{cond}	\\
|\Delta| &<& 0.25. \nonumber
\end{eqnarray}
These conditions guarantee that the higher order terms, which are evaluated around the critical point, are nowhere evaluated too far from the critical point such that the approximations would break down. 

We can now put the pieces together by using the general form of the optimal solution for $P^*(c)$ in Eq~(\ref{gensol}), together with the quartic ansatz for $P(g|c)$ in Eq~(\ref{pquartic}).   For each $c$, we evaluate  two entropies of the conditional distribution $P(g|c)$:
\begin{eqnarray}
S_2[P(g|c)] &=&\log_2\sqrt{2\pi e\sigma_g^2(c)} \label{ngauss} \\
S_4[P(g|c)]  &=& -\int dg \;P(g|c) \log_2 P(g|c). \label{nquartic}
\end{eqnarray}
$S_4$ is the noise entropy with higher-order terms included whenever conditions Eq (\ref{cond}) are met, and $S_2$ is the noise entropy in the Gaussian approximation.

Equation~(\ref{gensol}) can be rewritten in a numerically stable fashion by realizing that $P^*(c) |d\bar{g}/dc|^{-1}=P^*(\bar{g})$, that is, that the optimal distribution of mean output levels is given by
\begin{equation}
P^*(\bar{g})=2^{-S[P(g|\bar{g})]}/Z. \label{cap2}
\end{equation}
To join  the Gaussian and higher-order approximations consistently in the regimes away and near the critical point, the noise entropy in Eq~(\ref{cap2}) is chosen to be the pointwise minimum of $S_4(\bar{g})$ and $S_2(\bar{g})$. Finally, the information is again $I=\log_2 Z$, with 
\begin{equation}
Z=\int_0^{\bar{g}(C)} d\bar{g}\; 2^{-S[P(g|\bar{g})]}. \label{cap2}
\end{equation}

\subsection{Information transmission in the bistable regime}
\label{infobi}

We now discuss the information capacity in the bistsable regime, away from the critical line. In this regime, each value of the input $c$ can give rise to multiple solutions of the steady state equation, Eq~(\ref{ss}).  In the simplest case (which includes the MWC regulatory functions), there will be  two stable solutions, $\bar{g}_1(c)$ and $\bar{g}_2(c)$, and a third solution, $\bar{g}_3(c)$, that is unstable. In equilibrium, the system will be on the first branch with weight $w_1(c)$ and on the second with weight $w_2(c)$. Here we place upper bound on the information $I(c;g)$, again in the small noise approximation.  This will be useful since, as we will see, even this upper bound is always less than the information which can be transmitted in the monostable or critical regime, and so we will be able to conclude that the optimal parameters for which we are searching are never in the bistable regime.

In the bistable regime, the small noise approximation (again, away from the critical line) means that the conditional distributions are well approximated by a mixture of Gaussians,
\begin{eqnarray}
\label{pgcdef}
P(g|c) &=& w_1(c) \frac{1}{\sqrt{2 \pi  \sigma^2_{1}(c)}} e^{-\frac{(g-\bar{g}_1(c))^2}{2\sigma^2_{1}(c)}}
\nonumber\\
&&\,\,\,\,\,\,\,\,\,\, +w_2(c) \frac{1}{\sqrt{2 \pi  \sigma^2_{2}(c)}} e^{-\frac{(g-\bar{g}_2(c))^2}{2\sigma^2_{2}(c)}}  .\label{bigauss}
\end{eqnarray}
To compute the information we need two terms, the total entropy and the conditional entropy. The conditional entropy takes a simple form if we assume the noise is small enough that the Gaussians don't overlap.  Then a direct calculation shows that, as one might expect intuitively, the conditional entropy is just the weighted sum of the entropies of the Gaussian distributions, plus a term that reflects the uncertainty about which branch the system is on,
\begin{eqnarray}
S[P(g|c)] &=& {1\over 2}\sum_{{\rm i}=1}^2 w_{\rm i}(c) \log_2 \left[ 2\pi e \sigma_{\rm i}^2 (c)\right] 
\nonumber\\
&&\,\,\,\,\,\,\,\,\,\, - \sum_{{\rm i}=1}^2 w_{\rm i}(c) \log_2 w_{\rm i}(c) .
\end{eqnarray}

Implementing the small noise approximation for the total entropy is a bit more subtle. We have, as usual,
\begin{eqnarray}
P(g) &\equiv& \int dc\, P(c) P(g|c)\nonumber\\
& =&  \int dc\, P(c) {{w_1(c)}\over\sqrt{2\pi\sigma_1^2(c)}} \exp\left[ - {{(g - \bar g_1 (c))^2}\over{2\sigma_1^2(c)}}\right] 
\nonumber\\
&&\,\,\,\,\,\,\,\,\,\,+ 
\int dc\, P(c) {{w_2(c)}\over\sqrt{2\pi\sigma_2^2(c)}} \exp\left[ - {{(g - \bar g_2 (c))^2}\over{2\sigma_2^2(c)}} \right] .\nonumber\\
&&
\end{eqnarray}
If the noise is small, each of the two integrals is dominated by values of $c$ near the solution of the equation $g= g_{\rm i}(c)$; let's call these solutions $\hat c_{\rm i}(g)$.  Notice that these solutions might not exist over the full range of $g$, depending on the structure of the branches.  Nonetheless we can write
\begin{eqnarray}
P(g)& \approx& \left[ w_1 (c)P(c)  {\bigg |} {{d\bar g_1 (c)}\over {dc}} {\bigg |}^{-1}\right]_{c = \hat c_1(g)}\nonumber\\
&&\,\,\,\,\,\,\,\,\,\,+ \left[ w_2 (c)P(c)  {\bigg |} {{d\bar g_2 (c)}\over {dc}} {\bigg |}^{-1}\right]_{c = \hat c_2(g)} ,
\end{eqnarray}
with the convention that if we try to evaluate $w_{\rm i}(c)$ at a non--existent value of $\hat c_{\rm i}$, we get zero.    Thus, the full distribution $P(g)$ is also a mixture,
\begin{equation}
P(g) = f_1 P_1(g) + f_2 P_2(g).
\end{equation}
The fractional contributions of the two distributions are
\begin{equation}
f_{\rm i} = \int_{\rm i} dc\, P(c) w_{\rm i}(c)  ,
\end{equation}
where $\int_{\rm i} dc \, \cdots$ denotes an integral over the regions along the $c$ axis where the function $\hat c_{\rm i}(g)$ exists, and the (normalized) component distributions are 
\begin{equation}
P_{\rm i}(g)  = {1\over {f_{\rm i}}} \left[ w_{\rm i} (c)P(c)  {\bigg |} {{d\bar g_{\rm i} (c)}\over {dc}} {\bigg |}^{-1}\right]_{c = \hat c_{\rm i}(g)}
\end{equation}
The entropy of a mixture is always less than the average entropy of the components, so we have an upper bound
\begin{widetext}
\begin{eqnarray}
S[P(g)] &\leq& -\sum_{{\rm i}=1}^2 f_{\rm i}\int dg\, P_{\rm i}(g) \log_2 P_{\rm i}(g)\\
&=& -\sum_{{\rm i}=1}^2  \int dg\, \left[ w_{\rm i} (c)P(c)  {\bigg |} {{d\bar g_{\rm i} (c)}\over {dc}} {\bigg |}^{-1}\right]_{c = \hat c_{\rm i}(g)}\log_2  \left[ {1\over {f_{\rm i}}} w_{\rm i} (c)P(c)  {\bigg |} {{d\bar g_{\rm i} (c)}\over {dc}} {\bigg |}^{-1}\right]_{c = \hat c_{\rm i}(g)}\\
&=& -\sum_{{\rm i}=1}^2 \int_{\rm i} dc\, P(c) w_{\rm i}(c) \log_2  \left[ {1\over {f_{\rm i}}} w_{\rm i} (c)P(c)  {\bigg |} {{d\bar g_{\rm i} (c)}\over {dc}} {\bigg |}^{-1}\right]  .
\end{eqnarray}

An upper bound on the total entropy is useful because it allows us to bound the mutual information:
\begin{eqnarray}
I(c;g) &\equiv& S[P(g)] - \int dc\, P(c) S[P(g|c)]\\
&\leq& -\sum_{{\rm i}=1}^2 \int_{\rm i} dc\, P(c) w_{\rm i}(c) \log_2  \left[ {1\over {f_{\rm i}}} w_{\rm i} (c)P(c)  {\bigg |} {{d\bar g_{\rm i} (c)}\over {dc}} {\bigg |}^{-1}\right] 
\nonumber\\
&&\,\,\,\,\,\,\,\,\,\,-  {1\over 2}\sum_{{\rm i}=1}^2 \int dc\, P(c) w_{\rm i}(c) \log_2 \left[ 2\pi e \sigma_{\rm i}^2 (c)\right] +\sum_{{\rm i}=1}^2 \int dc\, P(c) w_{\rm i}(c) \log_2 w_{\rm i}(c)\\
&=& -\int dc\, P(c) \log_2 P(c) + {1\over 2}\int dc\, P(c) \sum_{{\rm i}=1}^2  w_{\rm i}(c) \log_2 \left[    {\bigg |} {{d\bar g_{\rm i} (c)}\over {dc}} {\bigg |}^2 {1\over{2\pi e \sigma_{\rm i}^2 (c)}}\right] - \left( -  \sum_{{\rm i}=1}^2 f_{\rm i} \log_2 f_{\rm i} \right)\\
&\leq&  -\int dc\, P(c) \log_2 P(c) + {1\over 2}\int dc\, P(c) \sum_{{\rm i}=1}^2  w_{\rm i}(c) \log_2 \left[    {\bigg |} {{d\bar g_{\rm i} (c)}\over {dc}} {\bigg |}^2 {1\over{2\pi e \sigma_{\rm i}^2 (c)}}\right] ,\label{Ibound_N}
\end{eqnarray}
\end{widetext}
where in the last step we use the positivity of the entropy associated with the mixture weights $\{f_{\rm i}\}$.

We can now ask for the probability distribution $P(c)$ that maximizes the upper bound on $I(c;g)$, and in this way we can bound the capacity of the system.  Happily, the way in which the bound depends on $P(c)$, in  Eq (\ref{Ibound_N}), is not so different from the dependencies that we have seen in the monostable case [Eq~(\ref{pointback})], so we can follow a parallel calculation to show that  
\begin{eqnarray}
 I(g;c) &\leq& \log_2 \left[\int_0^C dc \,e^{-\phi(c)}\right]\\
 \phi(c) &=&   \sum_i  w_i(c) \ln \left[\sqrt{2 \pi e \sigma^2_{g_i}(c)}  \left|\frac{d \bar{g}_i(c)}{d c} \right|^{-1} \right]. \label{multii}
\end{eqnarray}
Finally, to find the weights $w_{\rm i}(c)$ we can numerically integrate the Fokker--Planck solution in Eq~(\ref{intg}) to find
\begin{eqnarray}
w_1(c) &=& \int_0^{\bar{g}_3(c)} dg' \;P(g'|c), \\
w_2(c) &=& \int_{\bar{g}_3(c)}^{\infty} dg'\; P(g'|c). \label{weights}
\end{eqnarray}

To summarize, we have derived an upper bound on the information transmitted between the input and the output. The tightness of the bound is related to the applicability of the ``no overlap''  approximation, which for MWC--like regulatory functions should hold very well, as we have verified numerically.   If only one of the weights $w_{\rm i} \neq 0$, our results reduce to those in the monostable case, as they should.

\section{Results}
 
\begin{figure}[t] 
\centering
\includegraphics[width=\linewidth]{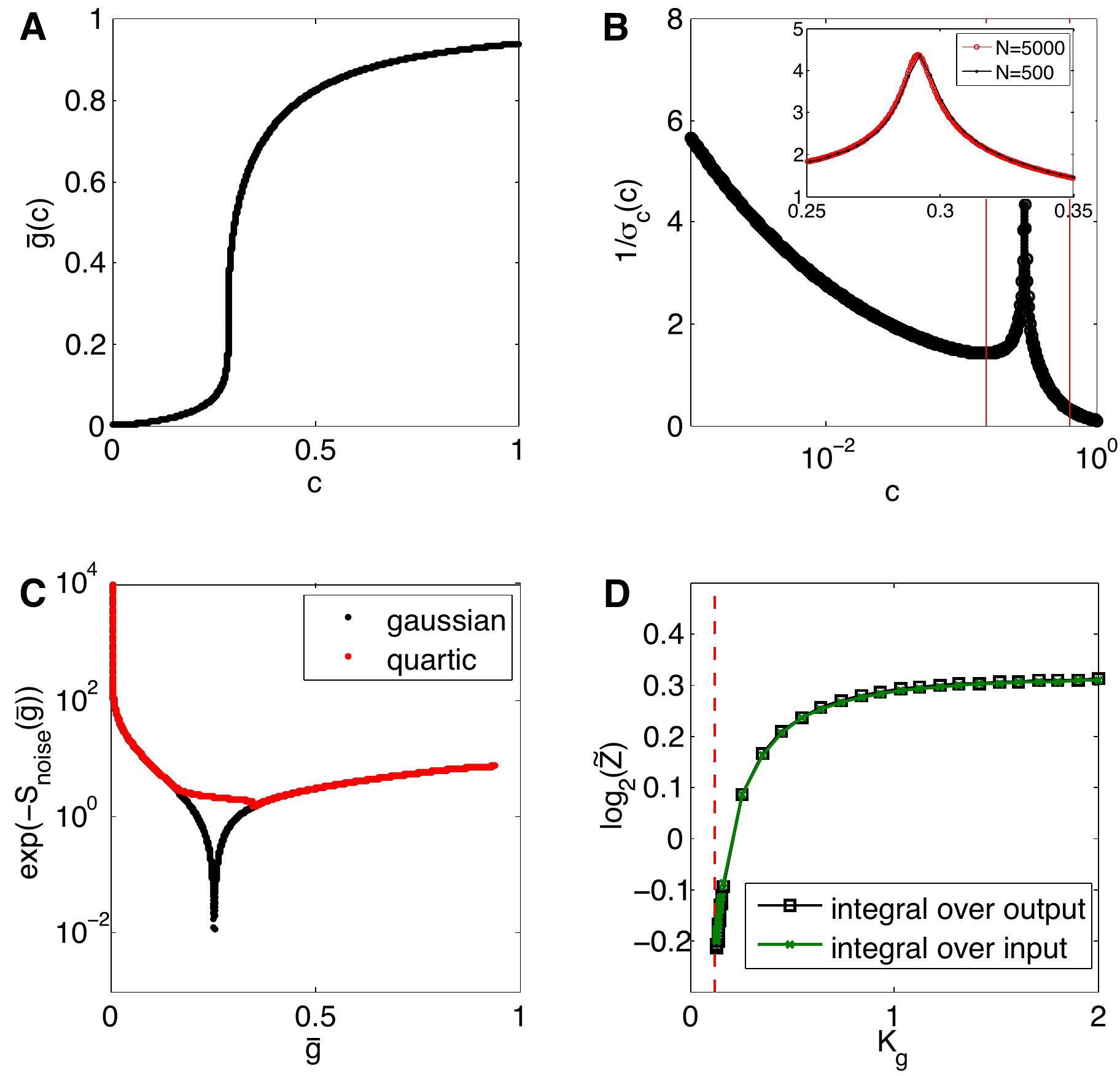} 
\caption{Computing information transmission close to the critical point. {\bf A)} An input/output relation $\bar{g}(c)$ for $n_g=2$, $C=1$ and $K_g=K_g^*(n_g)+0.004$, showing a self-activating gene with an almost critical value of $K_g$. {\bf B)} Noise in the input from Eq~(\ref{sigmac}), which is the integrand in the expression for information in Eq~(\ref{cap1}), shows an incipient divergence between red vertical bars. Inset: zoom-in of the peak shows that it can be sampled well by increasing the number of bins. Different plot symbols indicate domain discretization into 500 (black dots) and 5000 (red circles) bins. At the critical point the divergence is hard to control numerically. {\bf C)} An alternative way of computing the same information, by integrating in the output domain as in Eq~(\ref{cap2}). Shown is the integrand in the gaussian noise approximation [black, $S_2(\bar{g})$ from Eq~(\ref{ngauss})] and with quartic corrections [red, $S_4(\bar{g})$ from  Eq~(\ref{nquartic})]. At the critical point higher order corrections regularize the integrand, while away from the critical point the integrand smoothly joins with the gaussian approximation. This approach is stable numerically both away from and in the critical regime. {\bf D)} Information $I=\log_2 \tilde{Z}$ as a function of $K_g$ for $n_g=2$. Critical $K_g^*=1/8$ is denoted by a dashed red line. Integration across $\bar{g}$ in the output domain with quartic corrections (squares) agrees well with the integration across $c$ in the input domain (crosses) away from $K_g^*$, but also smoothly extends to the critical $K_g^*$.  This is a cut across the capacity plane in Fig~\ref{f-4}A (denoted by a dashed yellow line) for $C=1$. } 
\label{f-3}
\end{figure}

We begin by showing that the analytical calculations presented in the previous section can be  carried out numerically in a stable fashion, both away from and in the critical regime.   We recall  that the information transmission is determined by an integral $\tilde Z$ [Eq~(\ref{cap1})], and that because we are working in the small noise approximation we have a choice of evaluating this as an integral over the input concentration $c$ or an integral over the mean output concentrations $\bar g$.  Figure \ref{f-3} shows the behavior of the integrands in these two equivalent formulations when we have chosen parameters that are close to the critical point in a self--activating gene.  The key result is that, once we include terms beyond the Gaussian approximation to $P(g|c)$ following the discussion in Section \ref{critinfo}, we really do have control over the calculation, and find smooth results as the parameter values approach criticality.  Thus, we can compute confidently, and search for  parameters of the regulatory function ${\mathbf\theta}\equiv \{c_{1/2},K_c,n_c,K_g,n_g\}$ that maximize information transmission.

\begin{figure*}[t] 
\centering
\includegraphics[width=5in]{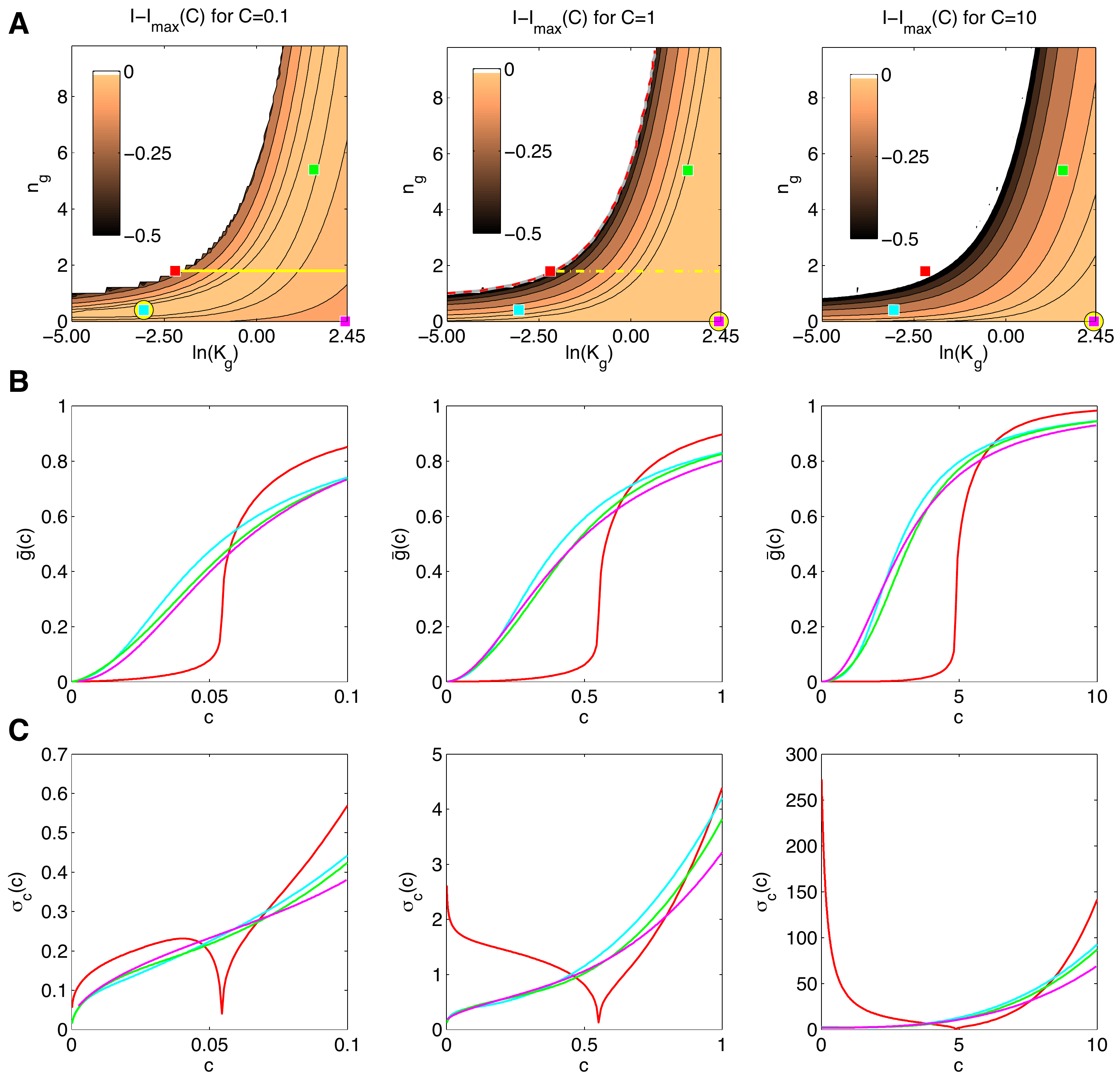} 
\caption{Information transmission as a function of self-activating parameters $\{n_g,K_g\}$ for three values of $C$; $C=0.1$ (left column), $C=1$ (middle column), $C=10$ (right column). For each value of $\{n_g,K_g\}$ the three remaining parameters $\{c_{1/2},n_c,K_c\}$ are optimized to maximize information. {\bf A)} The capacity planes showing the decrease in capacity in bits (colormap) from the maximal value achieved for an optimal choice of all parameters; the optimal set of parameters is denoted by a colored square on a yellow circle. The white upper left region of each capacity plane corresponds to the bistable region. For low $C=0.1$ the maximum is achieved in the interior of the domain (cyan square), while for $C=1,10$ $n_g$ is driven to 0 (magenta squares), corresponding to the non-self-interacting solution. Red square represents an example critical system at $n_g=2$ and the green square represents a self-interacting system with a high value for $n_g$. The yellow dashed line in $C=1$ plane represents a cut shown in detail in Fig~\ref{f-3}D. The yellow solid line in $C=0.1$ plane represents a cut shown in detail in Fig~\ref{f-5}A. The red dashed boundary of the critical region in $C=1$ plane is analyzed in detail in Figs~\ref{f-5}B, \ref{f-5}C. {\bf B)} Input/output relations for 4 example systems denoted by colored squares in A. Despite substantially different values for $\{n_g,K_g\}$, the optimization of remaining parameters makes the input/output curves look very similar to the optimal solutions, except for the critical (red) curves. {\bf C)} The effective input noise $\sigma_c(c)$ for the selected systems.   } 
\label{f-4}
\end{figure*}

We start the optimization by choosing the parameter values ${\mathbf\theta_g}\equiv \{n_g,K_g\}$ which describe the self--interaction term, and then holding these fixed while we optimize the remaining ones, ${\mathbf\theta_c}\equiv\{c_{1/2},n_c,K_c\}$. In all these optimizations the parameter $K_c$ is driven to zero, and in this limit the MWC regulatory function of Eq~(\ref{mwc}) simplifies to something more like the Hill function,
\begin{eqnarray}
f(g,c)&=&\frac{c^{n_c}}{c^{n_c}+ c_{1/2}^{n_c}e^{-\tilde{F}(g)}} \\
\tilde{F}(g)&=&n_g\ln\left( {{1+g/K_g}\over{1+1/(2K_g)}}\right)   .
\end{eqnarray}
Once we have optimized ${\mathbf\theta_c}$, we can explore the information capacity as a function of ${\mathbf\theta_g}$, at varying values of the remaining parameter in the problem, the maximal concentration $C$ of transcription factors.    Figure~\ref{f-4}A maps out the ``capacity planes,''  $I^*(n_g,K_g; C)$ at fixed  $C$.    In detail, we show $I^*(n_g,K_g;C)-I^*_{\rm max}(C)$ for three choices of our parameter $C$, where $I^*_{\rm max}(C)$ is the information obtained with the optimal choice of $n_g$ and $K_g$; the best choice of parameters is depicted as a yellow circle in the capacity plane. 

For large values of $C$, $C=1,10$, the optimal solution is at $n_g\rightarrow 0$ or $K_g\rightarrow \infty$ (magenta square in the lower right corner), which drives the self--activation term  in Eq~(\ref{mwc}) to zero, towards a noninteracting solution. We have checked that these solutions correspond to optimal solutions for a single noninteracting gene found in our previous work  \cite{wt2010}. As $C$ is decreased, however, the optimal combination $\{n_g,K_g\}$ shifts towards the left in the capacity plane (cyan square for $C=0.1$), exhibiting a shallow but distinct maximum in information transmission. If we examine the mean input/output relations in Fig~\ref{f-4}B, we find nothing dramatic: the critical (red) solutions seem to have lower capacities (which we carefully reexamine blow), while other quite distinct parameter choices for $\{n_g,K_g\}$ nevertheless generate very similar mean input/output relations, because of the freedom to optimally choose $\theta_c$ parameters. The behavior of effective noise in the input, $\sigma_c(c)$, given by Eq~(\ref{sigmac}) and shown in Fig~\ref{f-4}C, is more informative; recall that $\int dc\;\sigma^{-1}_c(c)$ is proportional to the information transmission.   Noninteracting (magenta) solutions always have the lowest amount of noise at high input concentrations ($c\sim C$). As the self--interaction turns on, the noise at high input increases, but that increase can be traded off for a decrease in noise at medium and low $c$. While for low $C=0.1$  the critical (red) solution is never optimal, the solution with some self--activation manages to deliver an additional $\sim 0.2$ bits of information. We have verified that for $10\times$ smaller value of $C=0.01$ the capacity plane is qualitatively the same, exhibiting the peak at a nontrivial (but still not critical) choice of $\{n_g\approx 0.51,K_g\approx 0.11\}$ (not shown).

Intuitively, the self-activation parameters $\theta_g$ have three direct effects on the information transmission: they change the shape of the input/output curve, the self--activation feeds some of the output noise back into the input, and the  time $\tau$ (protein lifetime) that averages the input noise component  gets renormalized to $\tau\rightarrow \tau  (1-\partial f/\partial g)^{-1}$. The changes in the mean input/output relation can be partially compensated for by the correlated changes in the $\theta_c$, as we observed in our optimal solutions, suggesting that regardless of the underlying microscopic parameters, it is the shape of $\bar{g}(c)$ itself that must be optimal.  The increase in averaging time acts to increase the information, thus favoring self--activation. However, this will  simultaneously increase the noise in the output that feeds back, as well as drive $\bar{g}(c)$ towards infinite steepness at criticality, restricting the dynamic range of the output. At low $C$ there is a parameter regime where increasing the integration time will help decrease the (dominant) input noise enough to result in a net gain of information. At high $C$, input noise is minimal  and thus this averaging effect loses its advantage; instead, feedback simply  acts to increase the total noise by reinjecting the output noise at the input, so that optimizing information transmission drives the self--interaction to zero.
 \begin{figure*}[t] 
\centering
\includegraphics[width=\linewidth]{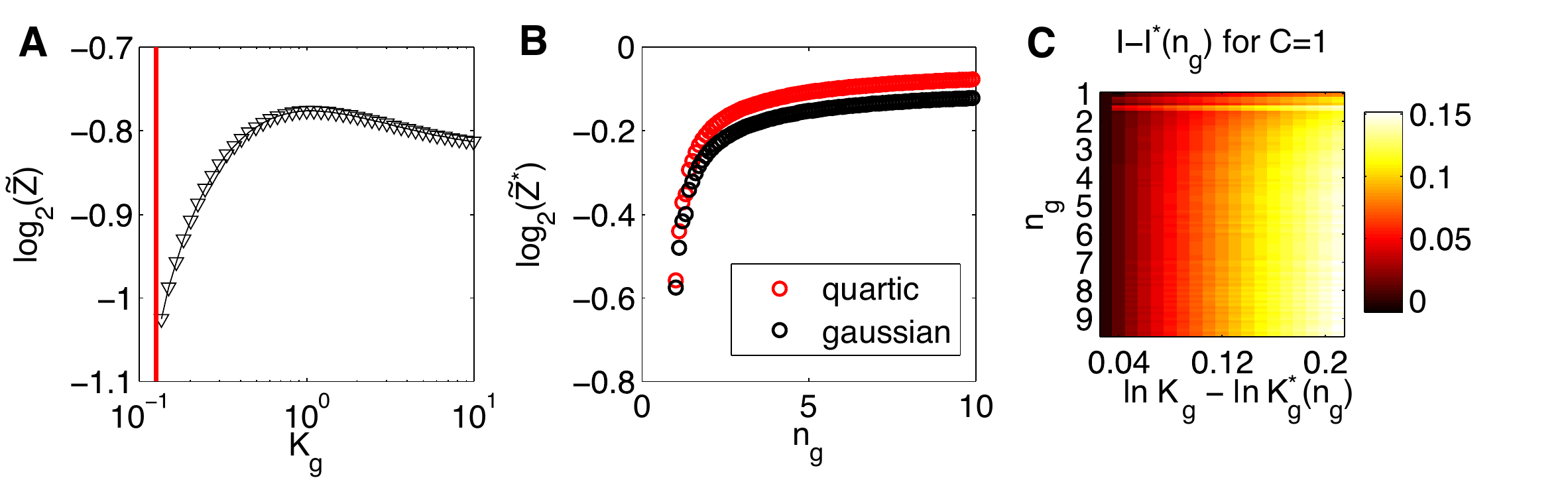} 
\caption{Information transmission close to the critical line. {\bf A)} A detailed scan of information for $C=0.1$ as the function of $K_g$ for $n_g=2$ and optimal choice for $\theta_c$ for every $K_g$ exhibits a peak for a nontrivial value of $K_g$. {\bf B)} The information transmission along the critical line for $C=1$ as a function of $n_g$ indicates that while including quartic corrections is important, the information at the critical line does not exhibit any large jumps (not shown). {\bf C)} A detailed scan of capacity close to the critical line for $C=1$ as a function of $n_g$ (vertical axis). Shown in color is the increase in information in bits compared to the value very close to the critical $I^*(n_g)$, as a function of the distance in $K_g$ from the critical value (horizontal axis). Across the range of $n_g$, going from the critical axis into the monostable domain increases the information.} 
\label{f-5}
\end{figure*}

Next we examine in detail the behavior of information transmission close to the critical region. Close to, but not at, the critical point we perform very fine discretization of the input range to evaluate the integral in Eq~(\ref{cap1}), as reported in Fig~\ref{f-3}B. To validate that the information indeed reaches a maximum at nontrivial values of $\{n_g,K_g\}$ when $C=0.1$, we cut through the capacity plane in Fig~\ref{f-4}A along the yellow line at $n_g=2$, and display the resulting capacity values in Fig~\ref{f-5}A (the results are numerically stable when integrated on $10^4$ or $10^3$ points). Unlike for $C=1$ and $C=10$, for $C=0.1$ the maximum is clearly achieved for a nontrivial value of $K_g$, but away from the critical line, confirming our previous observations. We further examine the capacity directly on the critical line, $K_g^*=(n_g-1)^2/(4n_g)$, as a function of $n_g$ at  $C=1$ (denoted in Fig~\ref{f-4}A with dashed red line). The capacity in this case can be calculated using Eq~(\ref{cap2}) and is shown in Fig~\ref{f-5}B. The capacity that includes quartic corrections is higher by $\sim 0.05-0.1$ bits than in the gaussian approximation, making the effect small but noticeable. We also confirmed that the capacity at the critical line joints smoothly with the capacity near the line, i.e. that there is no jump in capacity exactly at criticality, which presumably would be a sign of numerical errors. Figure~\ref{f-5}C finally validates that across the whole range of $n_g$ for $C=1$, small increases in $K_g$ above the critical value $K_g^*(n_g)$ always lead to an increase of information, demonstrating that the maximum is \emph{not} achieved on the critical line.

 \begin{figure}[b] 
\centering
\includegraphics[width=\linewidth]{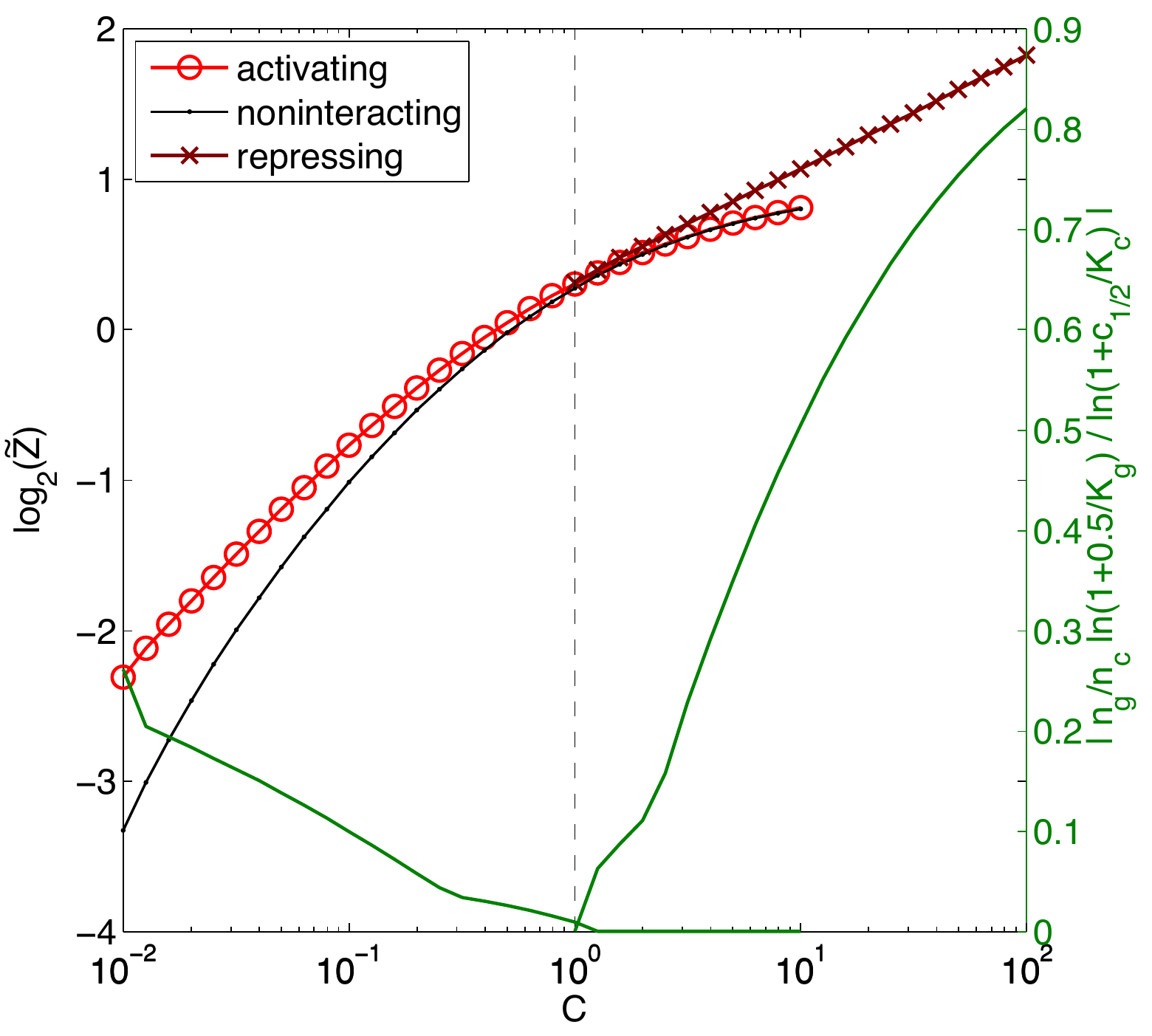} 
\caption{Information transmission in a system where all parameters $\theta$ are optimized as a function of the maximal input concentration $C$. The self-interacting system (red) allows for an arbitrary MWC-like regulatory function [Eq~(\ref{mwc})] with parameters $\theta=\{c_{1/2},n_c,K_c,n_g,K_g\}$. The noninteracting system (black) only has the MWC parameters $n_c,K_c$ and the leak $L$ (see Ref~\cite{wt2010}) which can be reexpressed in terms of $c_{1/2}$. Bright red line with circles shows self-activating solutions which are optimal for $C<1$, while dark red line with crosses shows self-repressing solutions, optimal for $C>1$. Plotted on the secondary vertical axis in green is the ratio between the self-interacting contribution to $F$, and the input contribution to $F$ in the expression for the MWC regulatory function [Eq~(\ref{mwc})]. For $C\sim 1$ where the interacting and noninteracting solution join, this term falls to 0, as expected. } 
\label{f-6}
\end{figure}

\begin{figure}[t] 
\centering
\includegraphics[width=\linewidth]{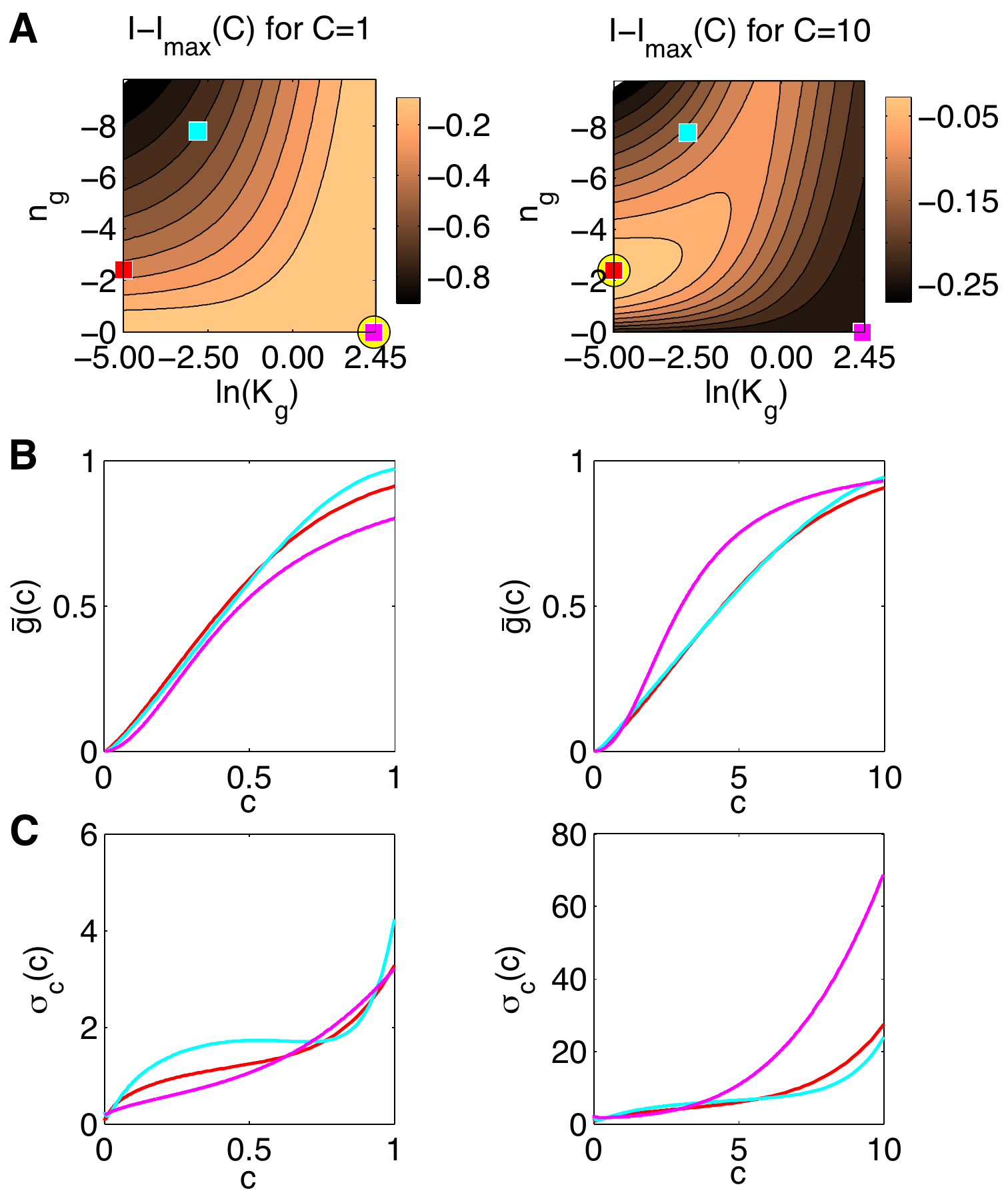} 
\caption{Information transmission as a function of self-repressing parameters $\{n_g,K_g\}$ for $C=1$ (left column) and $C=10$ (right column); plot conventions are the same as in Fig~\ref{f-4}. {\bf A)} The capacity decrease from the maximum value (achieved at the parameter choice indicated by a yellow circle) as a function of $\{n_g,K_g\}$. The maximum information transmission is achieved for a non-interacting case (magenta), for $C=1$. In contrast, for $C=10$, there is a non-trivial optimum for small values of $K_g$ and $n_g\approx-2.5$ (red). {\bf B)} The mean input/output solutions for three example systems from A (red, magenta, cyan).
{\bf C)} The effective noise in the input, $\sigma_c(c)$, for example solutions in A. 
} 
\label{f-7}
\end{figure}

We next turn to the joint optimization of all parameters  and plot the information transmission as a function of $C$ in Fig~\ref{f-6}. As we have discussed,  optimization  drives the strength of self--activation to zero for $C>1$ (but see below for self--repression), and at these high values of $C$ the result of full optimization coincides with the non--interacting case. As $C$ falls below one, the gain in information due to self--activation is increased, reaching a significant value of about a bit for $C=0.01$. 
As we have noted in Section \ref{MWC},  the self--activating effect of $g$ on its own expression can be changed into a self--repressing effect by simply flipping the sign of the parameter $n_g$.   To explore the optimization of such self--repressing genes, we thus optimized the parameters as before, now constraining $n_g\leq 0$.   Results in $\{ n_g, K_g\}$ plane are shown in Fig \ref{f-7}, for   $C=1$ and $C=10$.

We find that, for large $C$, the optimization process drives both $K_c$ and $K_g$ toward zero, so that  the effective input/output relation is given by 
\begin{equation}
f(g,c)=  {{c^{n_c}}\over {c^{n_c} + c_{1/2}^{n_c} \left({2g}\right)^{-n_g} }},
\end{equation}
with nonzero values of $n_g$ being optimal.  Why is self--repression optimal at large $C$, when self--activation is not?   Self--repression suppresses noise at high concentrations of the input (red vs magenta curves Fig~\ref{f-7}C for $C=10$) and allows the mean input/output curve to be more linear than in the non--interacting case (Fig~\ref{f-7}B), extending the dynamic range of the response.   Both these effects serve to increase information transmission.

It is remarkable that when we put together the self--activating and self--repressing solutions, we see that they join smoothly at $C=1$ (Fig \ref{f-6}):  self--activation is optimal for $C<1$, and self--repression is optimal for $C>1$, while precisely at $C=1$ the system that transmits the most information is non--interacting.  

All of this discussion has been in the limit where the maximal concentration of output molecules, $\gamma_{\rm max}$, is the same as the maximal concentration of input molecules, $C$,  so there is only one parameter that governs the structure of the optimal solution.  This makes sense, at least approximately, since both input and output molecules are transcription factors, presumably with similar costs, but  nonetheless  we would like to see what happens when we relax this assumption.  Intuitively, if we let $\gamma_{\rm max}$ become large, the system can achieve the advantages of feedback while the impact of noise being fed back into the system should be reduced.  

If we look at Eq (\ref{cap1}) for $\tilde Z$, which controls the information capacity, we can take the limit $\gamma_{\rm max}\rightarrow \infty$ to find
\begin{equation}
\tilde Z= \int_0^C dc \frac{\frac{\partial f}{\partial c}\left(1-\frac{\partial f}{\partial g}\right)^{-\frac{1}{2}}}{\sqrt{ \bar{g} + \left(\frac{\partial f}{\partial c}\right)^2c  }}. \label{cap_lim}
\end{equation}
Now the only place where feedback plays an explicit role is in the term $\left(1-\frac{\partial f}{\partial g}\right)^{-\frac{1}{2}}$, which comes from the lengthening of the integration time, which in turn serves to average out the noise in the system.  All other things being equal (which may be hard to arrange), this suggests that information transmission will be maximized if the system approaches the critical point, where $\partial f/\partial g \rightarrow 1$.  The difficulty is that the system can't stay at the critical point for all values of the input $c$, so there must be a tradeoff between lengthening the integration time and using the full dynamic range.

\begin{figure}[bht] 
\centering
\includegraphics[width=\linewidth]{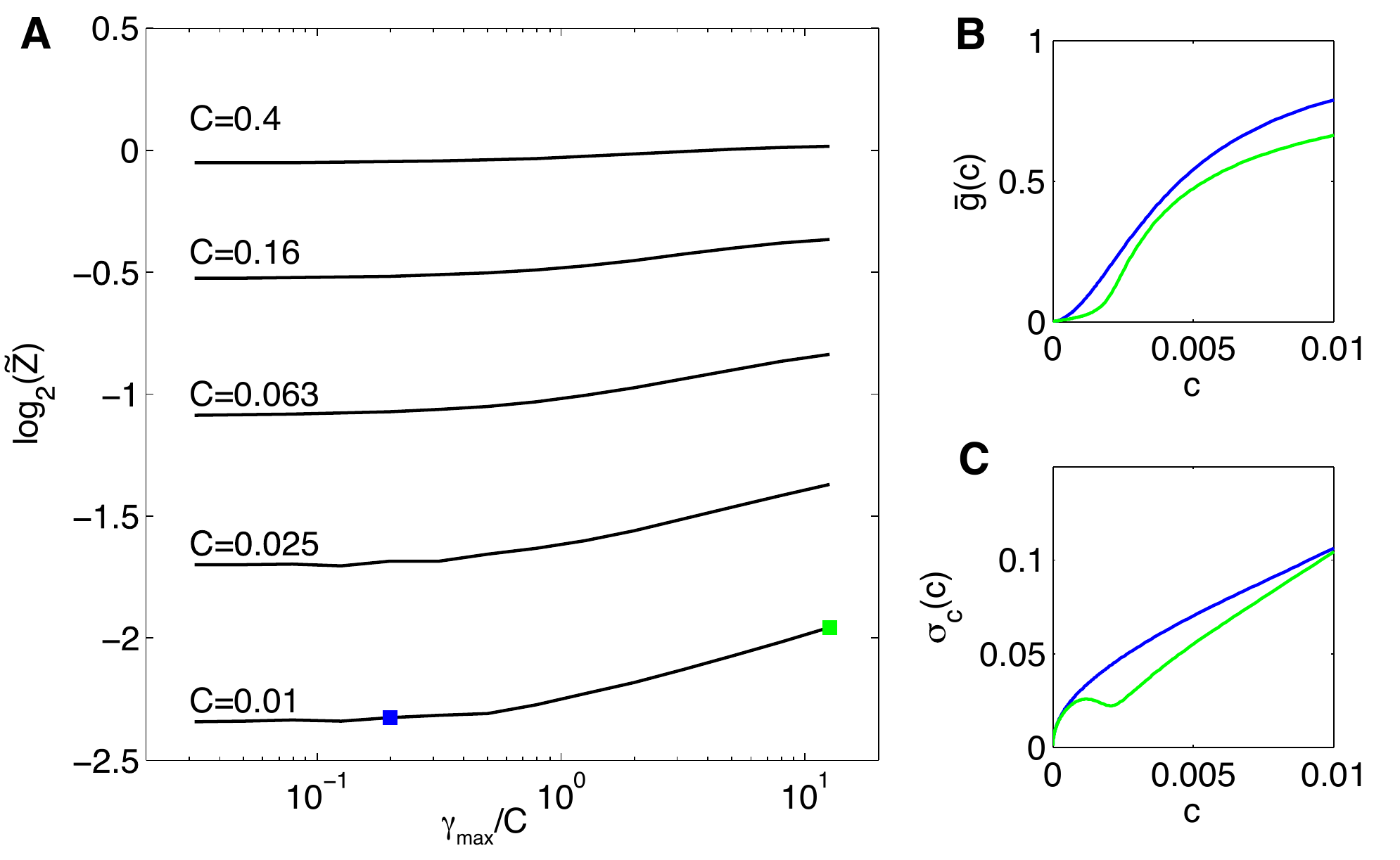} 
\caption{The dependence of information transmission in the self-activating case on the ratio $\gamma_{\rm max}/C$. {\bf A)} For various choices of $C$ indicated on the plot, the  information transmission with the optimal choice with respect to all parameters $\theta$ is shown as a function of $\gamma_{\rm max}/C$. Two special systems of interest (blue, green) are chosen for the lowest value of $C$. {\bf B)} The mean input/output relation for the blue and green system. The green system has a higher transmission, a steeper activation curve but a smaller dynamic range. {\bf C)} The effective noise in the input, $\sigma_c(c)$, for the blue and green systems. The green system is closer to critical at the point where the mean input/output curve has the highest curvature and the noise exhibits a dip.} 
\label{f-8}
\end{figure}

To explore more quantitatively, we treat $\gamma_{\rm max}/C$ as a parameter.   When $C$ is small, we know that self--activation is important, and in this regime we see from  Fig~\ref{f-8} that changing $\gamma_{\rm max}/C$  matters. On the other hand, for large values of $C$ we know that (at $\gamma_{\rm max}/C = 1$) optimization drives self--activation to zero, so we expect that there is less or no impact of allowing $\gamma_{\rm max}/C \neq 1$.  We also see that,   for a fixed small $C$, increasing $\gamma_{\rm max}/C$ drives the system closer towards the signatures of criticality---nonmonotonic behavior in the noise and a steepening of the input/output relation. In more detail, we can plot the value of $\mathrm{max}_c(\partial f/\partial g)$ as a function of $C$ and $\gamma_{\rm max}/C$, that is, check for each of the solutions in Fig~\ref{f-8}A how close the partial derivative $\partial f/\partial g$ comes to 1, which is a direct measure of criticality. We confirm that, for the simultaneous choice of small $C$ and large $\gamma_{\rm max}/C$,  we indeed have $\partial f/\partial g \rightarrow 1$.   In the extreme, if we choose $C=0.01$ and $\gamma_{\rm max}/C=10^4$,  we find that the  optimal $K_c$ and $K_g$ are driven towards small values (but since $c,C$ are small $K_c$ is not negligible); the optimal $n_g\approx 1.0662$. With this value of $n_g$, the corresponding critical value for $K_g$ would be $K_g^*(n_g)=0.001$, and the numerically found optimal value in our system is $K_g= 0.0012$. The critical value for $\bar{g}^*$ would be $\bar{g}^*(n_g)=0.031$, and indeed at this small value the optimal mean input/output relation has a strong kink, the effective noise $\sigma_c$ has a sharp dip and $\partial f/\partial_g$ at this point climbs to 0.9936. Numerically, therefore, we have all the expected indications of emerging criticality at very large $\gamma_{\rm max}$.   For less extreme values, we expect the optimum to result from the interplay between the input and transmitted noise contributions, which in general need not be on the critical line.

\begin{figure}[b] 
\centering
\includegraphics[width=\linewidth]{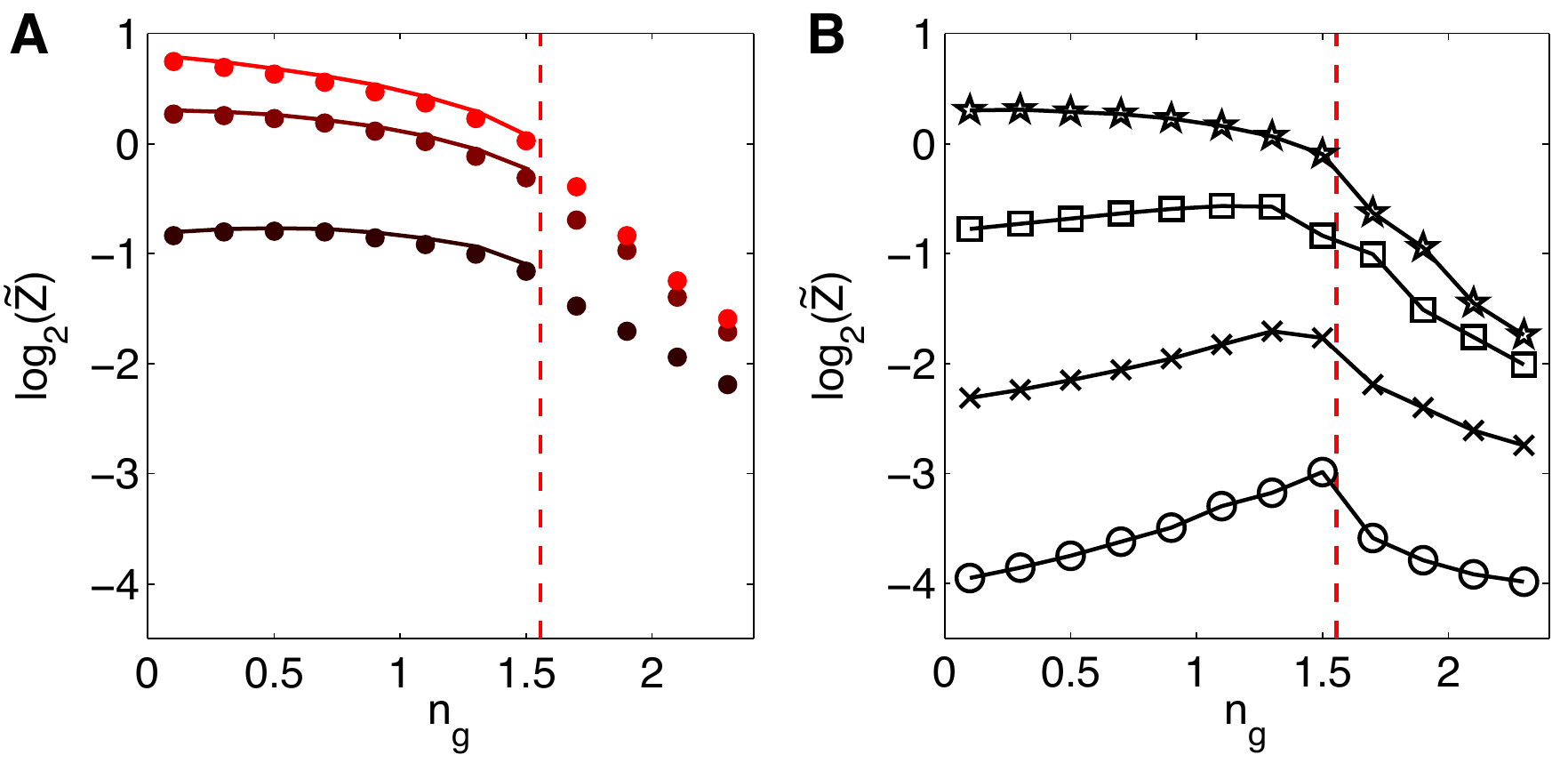} 
\caption{Crossing the critical line (dashed red) into the bistable region. {\bf A)} Capacity as a function of $n_g$ at fixed $K_g=0.05$, with the optimal choice of $\{n_c,K_c,c_{1/2}\}$ parameters, for three values of $C$ ($C=0.1,1,10$ dark to bright red, respectively). Dots show capacity calculation using the bistable code that can handle multiple branches using Eq~(\ref{multii}), solid line uses the monostable integration as in Eq~(\ref{cap1}). {\bf B)} Optimal capacity at very high  ratios $\gamma_{\rm max}/C=10^{3}$ for different value of $C$ ($C=10^{-3}, 10^{-2},10^{-1},1$: circles, crosses, squares, stars, respectively).  The optimum is pushed towards the critical line from the monostable side for $\gamma_{\rm max}/C$ large and $C$ small.  In all cases, information in the bistable regime is smaller than in the monostable regime.
} 
\label{f-9}
\end{figure}

To complete our exploration of the optimization problem, we have to consider parameter values for which the output has two locally stable values given a single input.  Quantitatively, in  the bistable regime we have to solve for both stable solutions $\bar{g}_{\rm i}(c)$, with ${\rm i}=1,2$, and for the unstable branch $\bar{g}_3(c)$. We can then evaluate the equilibrium probabilities $w_{\rm i}(c)$ of being on either of the stable branches using Eq~(\ref{weights}), and use Eq~(\ref{multii}) to compute the capacity. As shown in an example in Fig~\ref{f-9}A, we never find the optimal solutions in the bistable region---the capacity starts decreasing after crossing the critical line. Consistently with our argument that output and feedback noise must become negligible for the regime of small $C$ and large $\gamma_{\rm max}/C$, we find that optimization drives the system towards achieving maximal transmission  closer and closer the the critical line (which is approached from the monostable side), as shown in Fig~\ref{f-9}B.

\section{Discussion}

To summarize, we have analyzed in detail a single, self--interacting genetic regulatory element. As in previous work, we based our analysis on  three assumptions: (i) that the readout of the information $I(c;g)$ between the input and output happens in steady state,   (ii) that noise is small; and (iii) that the constraint limiting the information flow is the finite number of signaling molecules.   In addressing a system with feedback, assumption (ii) requires technical elaboration near the critical point, as discussed above.  But (i) requires a qualitatively new discussion for systems with feedback, because of the possibility of multistability. 
  
Our analysis, with the steady state assumption, shows that truly bistable systems do not maximize the information. Intuitively, this stems from the branch ambiguity:   for a given input concentration $c$ a bistable system can sit on either one of the stable branches with some probability, and this uncertainty contributes to the noise entropy, thereby reducing the transmitted information.  But reaching steady state involves waiting for two very different processes.  First, the system reaches a steady distribution of fluctuations in the neighborhood of each stable point, and then the populations of the two stable states equilibrate with one another.  As with Brownian motion in a double well potential (or a chemical reaction), these two processes can have time scales that are separated by an exponentially large factor.  

Alternatively, the timescales of real regulatory and readout processes could be such that the system does not have the time to equilibrate between the stable branches. In that case, the history (initial condition) of the system will matter, and the final value of the output $g$ will be determined jointly by the input $c$ and the past state of the output, $g_0$. Such regulatory elements can be very useful, because they retain memory of the past and are able to integrate it with the new input; a much studied biological example is that of a  toggle switch. The information measure we use here, $I(c;g)$, will not properly capture the abilities of such elements, unless we modify it to include the past state, e.g. into $I(\{c,g_0\};g)$: here both the input and current state together determine the output. Such computations are beyond the scope of this paper, but could make precise our intuitions about switches with memory. 

Multistability also allows for qualitatively new effects at higher noise levels. In our previous work we found that full information flow optimization (without assuming small noise) leads to higher capacities than a  small--noise calculation for an identical system and, moreover, that as noise grows, the optimal solutions start resembling a (noisy) binary switch where only the minimum and maximum states of input are populated in the optimal $P^*_{\rm in}(c)$ \cite{tkacik+al_08b}. At high noise, positive (even bistable) autoregulation could stabilize these two states and make them more distinguishable. 
In this case the design constraint for the genetic circuit is to use the smallest number of molecules that will prevent spontaneous flipping between the two branches on the relevant biological timescales \cite{billnips}.   In this limit regulatory elements can operate at high noise, with perhaps as few as tens of signaling molecules.

With these caveats in mind, our main results can be summarized as follows. Except at $C\sim 1$, the possibility of self--interaction always increases the capacity of genetic regulatory elements. For $C<1$, the optimal strategy is self--activation, while for $C>1$ it  is self--repression, as shown in Fig~\ref{f-6}. Self--repression allows the system to reduce the effective noise at high input levels and straighten the input/output relation, packing more ``distinguishable'' signaling levels into a fixed input range. Self--activation for small $C$ lengthens the effective integration time over which the (dominant) input noise contribution is averaged, thereby increasing information. The optimal level of self--activation is never so strong as to cause bistability, but does, for small $C$ and large $\gamma_{\rm max}/C$, push the optimal system towards the critical state.

An interesting observation about the nature of the optimal solutions is that self--activation which is strong enough to enhance information transmission may nonetheless not result in a functional input/output relation that looks very different from a system without self--activation, albeit with different parameters.  In such cases, information transmission is enhanced primarily by the longer integration time and reduced effective noise level.  This means that there need be no dramatic signature of self--activation, so that diagnosing this operating regime requires a detailed quantitative analysis. More generally, this result emphasizes that the same phenomenology can result from different parameter values, or even networks with different topology---in this case, with and without feedback.   

Stepping back from the detailed results, our goal in this paper was to make progress on understanding the optimization of information flow in systems with feedback by studying the simplest example.  The hope is that our results provide one building block for a theory of real genetic networks, on the hypothesis that they have been selected for maximizing information transmission.  As discussed in previous work \cite{tkacik+al_08a,tkacik+al_09,wt2010}, a natural target for such analysis is the well studied gap gene network in the early {\em Drosophila} embryo \cite{gapgenes}.  But we can also hope to connect with a broader range of examples.  The prevailing view of self--activation has been that its utility stems from the possibility of creating  a toggle (or a flip-flop) switch. This explanation, however, can only be true if self--activation is strong enough to actually push the system into the bistable regime. De Ronde and colleagues \cite{deronde} have improved on this intuition and have shown, in the linear response limit, that weak self--activation will increase the signal to noise ratio for dynamic signals, a function very different from the switch. Here we show that in the fully nonlinear, but steady state treatment,  monostable self--activation can be advantageous for information transmission. Furthermore, we show that there is a single control parameter, the ratio $C$ between the output and input noise strengths, which determines whether self--activation or self--repression is optimal. Since more and more quantitative expression data is available, especially for bacteria and yeast, one could try assessing how the use of both motifs correlates with the concentrations of input and output signaling molecules.

\begin{acknowledgments}
We thank T Gregor, EF Wieschaus, and especially CG Callan for helpful discussions.  Work at Princeton was supported in part by NSF Grants PHY--0957573 and CCF--0939370,  by NIH Grant R01 GM077599, and by the WM Keck Foundation.  For part of this work, GT was supported in part by NSF grant EF--0928048, and by the Vice Provost for Research at the University of Pennsylvania. 
 \end{acknowledgments}


\begin{thebibliography}{99}

\bibitem{shannon}
TM Cover \& JA Thomas JA (1991) \emph{Elements of Information Theory}. Wiley, New York.

\bibitem{inforeview}
G Tka\v{c}ik  \& AM Walczak (2011) Information transmission in genetic regulatory networks: a review. \emph{J Phys Condens Matt} {\bf 23:} 153102.

\bibitem{tkacik+al_08a}
G Tka\v{c}ik, CG Callan Jr \& W Bialek (2007) Information flow and optimization in transcriptional regulation.   {\em Proc Nat'l Acad Sci (USA)} {\bf 105:} 12265--12270.

\bibitem{tkacik+al_08b}
G Tka\v{c}ik, CG Callan Jr \& W Bialek (2008) Information capacity of genetic regulatory elements. {\em Phys Rev E} {\bf 78:} 011910.

\bibitem{ronen+al_02}
M Ronen, R Rosenberg, BI Shraiman \& U Alon (2002) Assigning numbers to the arrows: Parameterizing a gene regulation network by using accurate expression kinetics.  \emph{Proc Nat'l Acad Sci USA} {\bf 99:} 10555--10560.

\bibitem{tkacik+al_09}
G Tka\v{c}ik, A Walczak \& W Bialek (2009) Optimizing information flow in small genetic networks.  {\em Phys Rev E} {\bf 80:} 031920.

\bibitem{wt2010}
AM Walczak, G Tka\v{c}ik \& W Bialek (2010) Optimizing information flow in small genetic networks. II: Feed--forward interactions. \emph{Phys Rev E} {\bf 81:} 041905.

\bibitem{motifs}
SS Shen--Orr, R Milo, S Mangan \& U Alon  (2002) Network motifs in the transcriptional regulation network of Escherichia coli. \emph{Nat Genet} {\bf 31:} 64--8.

\bibitem{keseler}
IM Keseler, J Collado--Vides,  S Gama--Castro,  J  Ingreham, S Paley,  et al (2005) Ecocyc: a comprehensive database resource for Escherichia coli. \emph{Nucl Acids Res} {\bf 33:} D334--337.

\bibitem{hermsen}
R Hermsen, B Ursem \& PR ten Wolde (2010) Combinatorial gene regulation using auto-regulation. \emph{PLoS Comput Biol} {\bf 6:} e1000813.

\bibitem{bintu1} 
L Bintu, NE Buchler, HG  Garcia, U Gerland, T  Hwa, J Kondev  \& R Phillips  (2005) Transcriptional regulation by the numbers: models. \emph{Curr Opin Genet Dev} {\bf 15:} 116.

\bibitem{bintu2} 
L Bintu, NE Buchler, HG  Garcia, U Gerland, T  Hwa, J Kondev, T Kuhlman  \& R Phillips  (2005) Transcriptional regulation by the numbers: application. \emph{Curr Opin Genet Dev} {\bf 15:} 125.


\bibitem{rosenfeld}
N Rosenfeld, MB  Elowitz \& U Alon  (2002) Negative autoregulation speeds the response times of transcription networks. \emph{J Mol Biol} {\bf 323:} 785--93.

\bibitem{becksei}
A Becksei  \& L Serrano  (2000) Engineering stability in gene networks by autoregulation. \emph{Nature} {\bf 405:} 590--3.

\bibitem{becksei2}
A Becksei, B Seraphin  \& L Serrano  (2001) Positive feedback in eukaryotic gene networks: cell differentiation by graded to binary response conversion. \emph{EMBO J} {\bf 20:} 2528--2535.

\bibitem{buchler} 
NE Buchler, U Gerland  \& T Hwa  (2003) On schemes of combinatorial transcriptional logic. \emph{Proc Nat'l Acad Sci USA} {\bf 100:} 5136--41.


\bibitem{lambda}
A Arkin, J Ross  \& HH McAdams (1998) Stochastic kinetic analysis of developmental pathway bifurcation in phage $\lambda$-infected Escherichia coli cells. \emph{Genet} {\bf 149:} 1633.


\bibitem{billnips}
W Bialek  (2001) Stability and noise in biochemical switches, in \emph{Advances in Neural Information Processing Systems}, Leen TK, Dietterich TG \& Tresp V eds, pp. 103--109 (MIT Press, Cambridge, 2001).

\bibitem{gardner}
TS Gardner, CR Cantor  \& JJ Collins (2000) Construction of a genetic toggle switch in Escherichia coli. \emph{Nature} {\bf 403:} 339--42.

\bibitem{bcd}
T Gregor, DW Tank, EF Wieschaus  \& W Bialek  (2007) Probing the limits to positional information. \emph{Cell} {\bf 130:} 153.

\bibitem{erdmann}
T Erdmann, M Howard  \& PR ten Wolde  (2009) Role of spatial averaging in the precision of gene expression. \emph{Phys Rev Lett} {\bf 103:} 258101.

\bibitem{bialek+setayeshgar_05}
W Bialek \& S Setayeshgar (2005) Physical limits to biochemical signaling.   {\em Proc Nat'l Acad Sci (USA)} {\bf 102:} 10040--10045. 

\bibitem{sanchez_011}
A Sanchez, HG Garcia, D Jones, R Phillips \& J Kondev (2011)
Effect of promoter architecture on the cell-to-cell variability in gene expression. \emph{PLoS Comput Biol} {\bf 7:} e1001100.


\bibitem{kepler}
TB Kepler  \& TC Elston  (2001) Stochasticity in transcriptional regulation: origins, consequences and mathematical representations. \emph{Biophys J} {\bf 81:} 3116-3136.

\bibitem{thattai}
M Thattai  \& A van Oudenaarden (2001) Intrinsic noise in gene regulatory networks. \emph{Proc Nat'l Acad Sci USA} {\bf 98:} 8614--9. 

%
\bibitem{elowitz+al_02}
MB Elowitz, AJ Levine, ED Siggia \& PD Swain (2002)
Stochastic gene expression in a single cell.  {\em Science} {\bf 297} 1183--1186.
%
\bibitem{ozbudak+al_02}
E Ozbudak, M Thattai, I Kurtser, AD Grossman \& A van Oudenaarden (2002)
Regulation of noise in the expression  of a single gene.   {\em Nature Gen} {\bf  31:}
69--73.
%
\bibitem{blake+al_03}
 WJ Blake, M Kaern, CR Cantor \& JJ Collins (2003)
Noise in eukaryotic gene expression. {\em Nature} {\bf 422:} 633--637.
%
\bibitem{raser+oshea_04}
JM Raser \& EK O'Shea (2004)
Control of stochasticity in eukaryotic gene expression.
{\em Science} {\bf 304:} 1811--1814.
%
\bibitem{rosenfeld+al_05}
 N Rosenfeld, JW Young, U Alon, PS Swain \& MB Elowitz (2005)
 Gene regulation at the single cell level. {\em Science} {\bf 307:} 1962--1965.
%
\bibitem{pedraza+oudenaarden_05}
JM Pedraza \& A van Oudenaarden (2005)
Noise propagation in gene networks.
{\em Science} {\bf 307:} 1965--1969.

\bibitem{ido_05}
I Golding, J Paulsson, SM Zawilski \& EC Cox (2005)
Real--time kinetics of gene activity in individual bacteria. {\em Cell} {\bf 123:} 1025--36.

\bibitem{zenklusen_08}
D Zenklusen, DR Larson \& RH Singer (2008)
Single--RNA counting reveals alternative modes of gene expression in yeast. {\em Nat Struct Mol Bio} {\bf 15:} 1263--71.

\bibitem{ido_11}
LH So, A Ghosh, C Zong, LA Sep\'ulveda, R Segev \& I Golding (2011)
General properties of transcriptional time series in {\em  Escherichia coli}. {\em Nature Genet} {\bf 43:} 554-560.

\bibitem{ziv}
E Ziv,I  Nemenman  \& CH Wiggins (2007) Optimal signal processing in small stochastic biochemical networks. \emph{PLoS ONE} {\bf 2:} e1077.

\bibitem{deronde}
WH de Ronde, F Tostevin  \& PR ten Wolde  (2010) Effect of feedback on the fidelity of information transmission of time-varying signals. \emph{Phys Rev E} {\bf 82:} 031914.

\bibitem{gardiner}
C Gardiner, {\em Stochastic Methods: A Handbook for the Natural and Social Sciences, 4th ed.} (Springer, 2009).

\bibitem{tkacik+bialek_07}
G Tka\v{c}ik \& W Bialek (2007)
Diffusion, dimensionality and noise in transcriptional regulation.  {\em Phys Rev E} {\bf 79:} 051901.

\bibitem{tkacik+al_08c} 
G Tka\v{c}ik, T Gregor \& W Bialek (2008) The role of input noise in transcriptional regulation. {\em PLoS One} {\bf 3:} e2774.

\bibitem{hill}
AV Hill (1910) The possible effects of the aggregation of the molecules of haemoglobin on its dissociation curves. \emph{J Physiol} {\bf 40}: Suppl iv-vii.

\bibitem{mwc}
J Monod, J Wyman \& JP Changeaux (1965) On the nature of allosteric transitions: a plausible model. \emph{J Mol Biol} {\bf 12:} 88--118.

\bibitem{mirny}
LA Mirny  (2010) Nucleosome-mediated cooperativity between transcription factors. \emph{Proc Nat'l Acad Sci USA} {\bf 107:} 22534--9.

\bibitem{gapgenes}
J Jaeger  (2011) The gap gene network. \emph{Cell Mol Life Sci} {\bf 68:} 243--74.



\end{thebibliography}
\end{document}